\documentstyle[12pt]{article}

\begin{document}

\setlength{\baselineskip}{0.30in}

\begin{flushright}
 INS-rep.-1087 \\
 KUCP-0077 \\
 January 1995
\end{flushright}

\begin{center}
\begin{Large}
  {\bf Quantum and Classical Aspects of }\\[3mm]
  {\bf Deformed $c=1$ Strings}
\end{Large}

\vspace{25pt}

\noindent
 T.NAKATSU,\raisebox{2mm}{{\footnotesize 1}}
 K.TAKASAKI \raisebox{2mm}{{\footnotesize 2}{$\dagger$}}
and
 S.TSUJIMARU \raisebox{2mm}{{\footnotesize 1}}

\vspace{18pt}

\begin{small}
$~^{1}${\it Institute for Nuclear Study, University of Tokyo}\\
       {\it Midori-cho, Tanashi, Tokyo 188, Japan}\\
$~^{2}${\it Department of Fundamental Sciences }\\
       {\it Faculty of Integrated Human Studies, Kyoto University}\\
       {\it Yoshida-Nihonmatsu-cho, Sakyo-ku, Kyoto 606, Japan}

\end{small}

\vspace{25pt}

\underline{ABSTRACT}

\end{center}

\vspace{10pt}

\begin{small}
%A nonperturbative treatment of $c=1$ strings in the
%black-hole background is developed on the basis of a
%matrix model approach recently proposed by Jevicki and
%Yoneya. The large $N$ limit becomes a nonrelativistic
%fermion system on a half line with the one-body potential
%$V(x)=-x^2/2+m/(2x^2)$.
%A spectral generating algebra isomorphic to
%$su(1,1)$ is constructed, and eigenfunctions of the
%one-body problem are given by the Whittaker functions.
%The associated scattering problem is also exactly solvable,
%and asymptotic fields
%(in the second-quantized system) and their $S$-matrix are
%explicitly determined.
%A detailed analysis of this problem shows that the asymptotic
%theory is essentially a two-dimensional relativistic
%free fermion theory.
%In the compactification at self-dual radius, the $S$-matrix
%between the in-coming and out-going asymptotic Hilbert
%spaces is represented by a $GL(\infty)$ element $g$.
%$g$ also plays the role of an intertwining operator of
%two underlying $W_{1+\infty}$ algebras.
%Polchinski's scattering equations in his collective
%field approach is reproduced from this intertwining relation
%as a ``classical limit'', $\hbar \rightarrow 0$,
%for a suitable choice of $\hbar \sim g_{st}$.
%These results are further translated into the language of
%the Toda lattice hierarchy. The intertwining relation
%of two $W_{1+\infty}$ algebras are used to derive string
%equations. Polchinski's scattering equations are shown to be
%exactly the classical (or ``dispersionless'') limit
%of these nonperturbative string equations.

The quantum and classical aspects of a
deformed $c=1$ matrix model proposed by
Jevicki and Yoneya are studied.
String equations are formulated in
the framework of the Toda lattice hierarchy.
The Whittaker functions now play the role of
generalized Airy functions in $c<1$ strings.
This matrix model has two distinct parameters.
Identification of the string coupling
constant is thereby not unique,
and leads to several different perturbative
interpretations of this model as a string theory.
Two such possible interpretations  are examined.
In both cases, the classical limit of the
string equations,
which turns out to give a formal solution
of Polchinski's scattering equations,
shows that  the classical scattering amplitudes
of massless tachyons are insensitive
to deformations of the parameters in the matrix model.

\end{small}
\vfill
\hrule
\vskip 3mm

\begin{small}

\noindent{$\dagger$}
Supported in part by the Grants-in-Aid for Scientific Research, the
Ministry of Education, Science and Culture, Japan.

\end{small}

%%%%%%%%%%%%%%%%%%%%%%%%%%%%%%%%%%%%%%%%%%%%%%%%%%%%%%%%%%%%%%%%%%%%%%%
\newcommand{\beqn}{\begin{equation}}
\newcommand{\eeqn}{\end{equation}}
\newcommand{\beqnarray}{\begin{eqnarray}}
\newcommand{\eeqnarray}{\end{eqnarray}}
\newcommand{\rd}{\partial}
\newcommand{\dfrac}[2]{ \frac{\displaystyle #1}{\displaystyle #2} }
\newcommand{\binom}[2]{ {#1 \choose #2} }
\newcommand{\res}{\;\mathop{\mbox{\rm res}}}
\newcommand{\diag}{\mathop{\mbox{\rm diag}}}
\newcommand{\cA}{{\cal A}}
\newcommand{\bA}{{\bf A}}
\newcommand{\Abar}{{\bar{A}}}
\newcommand{\cAbar}{{\bar{\cA}}}
\newcommand{\bAbar}{{\bar{\bA}}}
\newcommand{\cB}{{\cal B}}
\newcommand{\bB}{{\bf B}}
\newcommand{\Bbar}{{\bar{B}}}
\newcommand{\cBbar}{{\bar{\cB}}}
\newcommand{\bBbar}{{\bar{\bB}}}
\newcommand{\Hbar}{{\bar{H}}}
\newcommand{\cL}{{\cal L}}
\newcommand{\bL}{{\bf L}}
\newcommand{\Lbar}{{\bar{L}}}
\newcommand{\cLbar}{{\bar{\cL}}}
\newcommand{\bLbar}{{\bar{\bL}}}
\newcommand{\cM}{{\cal M}}
\newcommand{\bM}{{\bf M}}
\newcommand{\Mbar}{{\bar{M}}}
\newcommand{\cMbar}{{\bar{\cM}}}
\newcommand{\bMbar}{{\bar{\bM}}}
\newcommand{\bU}{{\bf U}}
\newcommand{\cW}{{\cal W}}
\newcommand{\bW}{{\bf W}}
\newcommand{\Wbar}{{\bar{W}}}
\newcommand{\Xbar}{{\bar{X}}}
\newcommand{\cWbar}{{\bar{\cW}}}
\newcommand{\bWbar}{{\bar{\bW}}}
\newcommand{\abar}{{\bar{a}}}
\newcommand{\nbar}{{\bar{n}}}
\newcommand{\tbar}{{\bar{t}}}
\newcommand{\ubar}{{\bar{u}}}
\newcommand{\utilde}{\tilde{u}}
\newcommand{\vbar}{{\bar{v}}}
\newcommand{\wbar}{{\bar{w}}}
\newcommand{\phibar}{{\bar{\phi}}}
\newcommand{\Psibar}{{\bar{\Psi}}}
%%%%%%%%%%%%%%%%%%%%%%%%%%%%%%%%%%%%%%%%%%%%%%%%%%%%%%%%%%%%%%%
%%%%%%%%%%%%%%%%%%%%%%%%%%%%%%%%%%%%%%%%%%%%%%%%%%%%%%%%%%%%%%%%

\newpage

\section{Introduction and Summary }

        Matrix models have been a very powerful tool for our understanding of
nonperturbative string theories in two or fewer space-time dimensions
\cite{GM}. If a matrix model is to be taken as a nonperturbative definition
of string theory, all solutions of string theory should be described by this
approach. In particular it is very interesting to provide a matrix model
formulation of the two-dimensional critical string theory in the black-hole
background \cite{BH1}, \cite{BH2} since it will give  us a full quantum
mechanical (and  possibly nonperturbative) understanding of
black-hole physics.
An important step in this direction has been taken by Jevicki and Yoneya
\cite{JY}.
They proposed  that a  stationary black-hole of mass $m$ with tachyon
condensation is described by the large $N$ hermitian matrix quantum mechanics
with potential
$V(\Phi)= \mbox{Tr} \left\{ -\Phi^2/2+ m\Phi^{-2}/2N^2 \right\}$.
This matrix model has two entirely different parameters,
the Fermi energy $\mu$  and the black-hole mass $m$.
They parametrize deformations of the linear dilaton solution
of the critical string theory
in a flat space-time background
(i.e. $\mu=m=0$).
It is also argued in \cite{JY} that,
whereas the string coupling constant in the standard
$c=1$ string is given by  $g_{st}\sim 1/\mu$,
that of the black-hole should be modified as
$g_{st}\sim 1/\sqrt{m}$.
These two different identifications
of the string coupling constant lead us to two different perturbative
string theoretical interpretations  of this deformed matrix model, i.e.,
a deformation from
a flat space-time background  by the black-hole
mass operator, and  a deformation from the black-hole background
by the Fermi level (both with tachyon condensates).
Many attempts have been done towards nonperturbative understanding of
this deformed $c=1$ string model.
Classical scattering amplitudes of massless tachyons
in the black-hole background  (i.e. $\mu=0$)
have been studied in \cite{JY},\cite{DR},\cite{DK},\cite{DAN}
mainly by  the
collective field approach to this deformed matrix model.

In this article,  we present a nonperturbative treatment of this deformed
matrix model.
The goal is to derive string equations that encode the dynamics of deformed
$c=1$ strings.
The tachyon scattering amplitudes will then be reproduced in the classical
limit
as the string coupling constant $g_{st}\sim \hbar \rightarrow 0$.
To formulate such string equations, we need to specify an underlying
integrable structure.

%        In this article we present a  nonperturbative treatment of
%this deformed matrix model. In particular,
%string equations for
%these deformed $c=1$ strings  are formulated
%and their origin are investigated.
%The tachyon scattering amplitudes can be reproduced from classical
%limit of these string equations.

            In the case of $c<1$ string theory
the nonperturbative partition function
of $(p,q)$ string is a $\tau$ function of the Kadomtsev-Petriashivil
(KP) hierarchy
specified by the  string equation \cite{Douglas} :
$\mbox{$[$}P,Q \mbox{$]$}=1$
where $P$ and $Q$ are differential operators with
$\mbox{deg}P=p$ and $\mbox{deg}Q=q$.
In the particular case of  $(p,1)$ string
\cite{AM},
the solution of the string equation is written
in terms of the generalized Airy equation
\begin{eqnarray*}
(\partial_{\lambda}^p-\lambda)f_0(\lambda)~=~0~~~,
\end{eqnarray*}
or equivalently its first order form
\begin{eqnarray*}
\lambda f_k(\lambda) ~=~ f_{k+p}(\lambda)-kf_{k-1}(\lambda)~~,~~
\partial_{\lambda}f_{k}(\lambda) ~=~ f_{k+1}(\lambda)~~~,
\end{eqnarray*}
where $0 \leq k \leq p-1$.
The partition function of this model is given by a
generalized matrix Airy function \cite{Kontsevich},\cite{Morozov} and
several topological aspects appear through the asymptotic expansion
of this function \cite{Kontsevich},\cite{NKNT}.
This matrix integral representation of $(p,1)$ string gives a
nonperturbative formulation of $c<1$ string theory.

               Is there  a counterpart of the generalized Airy
functions for these deformed $c=1$ strings?  Motivated by this
question we devote Section 2 to a  detailed study of a
nonrelativistic
fermion system which describes \cite{SW},\cite{GK},\cite{MPR}
the large $N$ limit of this deformed
matrix model.
Our fundamental standpoint is the observation that
the Whittaker functions give a $c=1$ counterpart
of the generalized Airy functions in $(p,1)$ strings.
We construct
two sets of operators $B_n$ and $\bar{B}_n$
($n \in N$) in this fermion system, part of which
turn out to be  creation operators
of massless tachyons in these deformed $c=1$ strings.
(This point is discussed at the end of Section 3
in comparison with  classical scattering data obtained from
the collective field approach.)
On the basis of the analysis of the one-body system
we  also introduce the asymptotic
fields for these nonrelativistic fermions.
Besides the $S$-matrix elements between these asymptotic fields,
the asymptotic forms of $B_n$ and $\bar{B}_n$  are
thus also  determined.
It should be stressed  that these results  are
derived essentially from
the recursion relations and
connection formula of the Whittaker functions $M_{\kappa,\mu}$
($W_{\kappa,\mu}$)
\begin{eqnarray*}
x\partial_xM_{\kappa,\mu}(x)&=&
\pm (x/2-\kappa)M_{\kappa, \mu}(x)+(\mu \pm \kappa+1/2)
M_{\kappa \pm 1,\mu}(x)~~~, \\
W_{\kappa,\mu}(x) &=&
\frac{\Gamma(-2 \mu)}{\Gamma(1/2-\mu-\kappa)}M_{\kappa, \mu}(x) +
\frac{\Gamma(2 \mu)}{\Gamma(1/2+\mu-\kappa)}M_{\kappa, -\mu}(x)~~~.
\end{eqnarray*}
As we show  in Section 4,
the asymptotic expressions
of $B_n$ and $\bar{B}_n$ determine string equations.

           In Section 3,
we consider the compactification at self-dual radius.
This makes the ingredients of Section 2 more tractable.
We identify the asymptotic Hilbert spaces of the nonrelativistic
fermion system with that of two-dimensional relativistic fermions.
The asymptotic operators
$B_n^{in,out}$ and $\bar{B}_n^{in,out}$ are then identified
with even generators of the $U(1)$ current subalgebra in
a fermionic realization of $W_{1+\infty}$ algebra obtained
from relativistic fermions.
For an example,
$B_n^{out}\sim g^{-1}W_{2n}^{(0)}g$ and
$\bar{B}_n^{out}\sim W_{-2n}^{(0)}$
are the standard generators of the
$U(1)$ subalgebra, and $g$ is an element of
$GL(\infty)$ that acts on the Hilbert space of
relativistic fermions.
Thus $g$ is exactly a counterpart of the
$S$-matrix of Section 2, now realized on the
Hilbert space of relativistic fermions.
Furthermore, as one can see from the aforementioned
expression of
$B_n^{in,out}$ and $\bar{B}_n^{in,out}$,
$g$ also plays the role of an intertwining operator for
$W_{1+\infty}$ algebra.
We shall indeed make use of this intertwining property of
$g$ to construct $W_{\infty}$ symmetry of the deformed
$c=1$ strings.

%       In Section 3 we consider the compactification at  self-dual radius.
%This makes the ingredients of Section 2 more tractable.
%By identifying the asymptotic
%Hilbert spaces of the nonrelativistic fermion system
%with that of two-dimensional relativistic fermions,
%the scattering realtions between the asymptotic operators
%$B_n^{in,out}$ (or $\bar{B}_n^{in,out}$) turn out to be
%described by the adjoint action of  a $GL(\infty)$ element
%$g$ on even generators $W_{2n}^{(0)}$
%(or $W_{-2n}^{(0)}$ ) in
%this  relativistic fermion system.
%For an example,
%$B_n^{out} (\sim g^{-1}W_{2n}^{(0)}g) $ and
%$\bar{B}_{n}^{out} (\sim W_{-2n}^{(0)})$.
%$g$ is exactly the $S$-matrix obtained in Section 2.
%By examining the adjoint action of $g$,
%it is also shown that
%these asymptotic Hilbert spaces have $W_{\infty}$ symmetry,
%whose generators are given by a linear combination of
%$W_{1+\infty}$ algebra  and their images under
%the adjoint action of $g$.
%It means that,
%though there are  two
%$W_{1+\infty}$ symmetries  (realized by the free relativistic fermions)
%in both the  $in$- and $out-$Hilbert spaces,
%they are not independent (linked by the intertwining operator $g$)
%and give rise to
%the $W_{\infty}$ symmetry of these deformed $c=1$ strings.

                        In the latter half of this section
we consider classical limits of these intertwining equations.
One can obtain a  classical counterpart of the intertwining equations
by identifying the Planck constant $\hbar$ with a particular combination
of the parameters ($\mu,m$)
of the system and taking $\hbar \rightarrow 0$ limit.
In the case of $\hbar=-1/(i \sqrt{m})$,  the classical limit  becomes
independent of $\mu$,  and gives a formal solution of
Polchinski's classical scattering equations
\begin{eqnarray*}
\alpha_{\pm}(y)=\alpha_{\mp}(y\mp 1/2 \ln (1+\alpha_{\pm}^2(y)))~~~,
\end{eqnarray*}
which are derived  in \cite{JY} by  the collective
field  approach to this deformed matrix model.
On the other hand, in the case of $\hbar=-1/(i\mu)$,
the classical limit becomes  independent of $m$,
and gives a  formal solution of another variation of
Polchinski's classical
scattering equations
\begin{eqnarray*}
\alpha_{\pm}(y)=\alpha_{\mp}(y\mp \ln (1+\alpha_{\pm}(y)))~~~,
\end{eqnarray*}
which are given  in \cite{P},\cite{MP} by the collective
field approach to the standard $c=1$ matrix model.

                 In Section 4,  string equations  are presented for these
$c=1$ strings (compactified at  self-dual radius).
As in the case of the standard $c=1$ string \cite{DMP},\cite{N},
\cite{HOP-Takasaki-EK-BX}, these string equations are formulated
in the framework of the Toda lattice hierarchy
\cite{bib:TT-review}, \cite{bib:UT-Toda}.
We begin Section 4 by a rather detailed overview on the Toda lattice
hierarchy in three different languages, that is, difference operators,
infinite matrices and free fermions.
A fundamental observation here is that,
though the space  of solutions of the Toda lattice hierarchy has a
$W_{1+\infty} \oplus W_{1+\infty}$ symmetry,
an intertwining relation as described in Section 3
can induce a relation between these two
$W_{1+\infty}$ symmetries.
This indeed occurs if the solution is characterized by
a fixed point condition under these symmetries,
and string equations are nothing else
than such a fixed point condition.
We then return to the deformed $c=1$ strings.
String equations turn out to have the forms
\begin{eqnarray*}
\bar{L}^{-2}&=&
\frac{1}{\mu^2+m}
\left(-(ML^{-1}-i\mu L^{-1})^2+mL^{-2} \right)~~~, \\
L^2 &=&
\frac{1}{\mu^2+m}
\left(-(\bar{M}\bar{L}+(1-i\mu)\bar{L})^2+m\bar{L}^2 \right)~~,
\end{eqnarray*}
where $L,M,\bar{L}$ and $\bar{M}$ are the Lax and Orlov-Shulman operators
of the Toda lattice hierarchy.
The solution of these string equations, as in the case of $m=0$ \cite{DMP},
is given by the generating function of
(nonperturbative) tachyon scattering amplitudes
(expressed in terms of the relativistic fermions)
\begin{eqnarray*}
e^{\mbox{$\cal{F}$}(t,\bar{t})}=
<0|e^{\sum_{k \geq 1}t_kW_k^{(0)}} g
e^{-\sum_{k \geq 1}\bar{t}_kW_{-k}^{(0)}}|0>  .
\end{eqnarray*}
The string coupling constant $g_{st}(\sim \hbar)$ will
be recovered by the rescaling of parameters,
$t_k \rightarrow t_k/\hbar$ and
$\bar{t}_k \rightarrow \bar{t}_k/\hbar$.
The results of Section 3 for clasical limit
are now described in the terminology
of the dispersionless Toda hierarchy. The classical string equations
are
\begin{eqnarray*}
\bar{\cL}^{-2}~=~\cM \cL^{-2}+\cL^{-2}~,~
\cL^2~=~\bar{\cM}^2\bar{\cL}^2+\bar{\cL}^2~~,
\end{eqnarray*}
for the case of $\hbar = -1/(i\sqrt{m})$,
\begin{eqnarray*}
\bar{\cL}^{-2}~=~(\cM \cL^{-1}+\cL^{-1})^2~,~
\cL^2 ~=~ (\bar{\cM}\bar{\cL}+\bar{\cL})^2~~,
\end{eqnarray*}
for the case of $\hbar = -1/(i \mu)$.
Here $\cL,\cM,$ etc are the classical analogues of the Lax and
Orlov-Shulman operators of the Toda lattice hierarchy.
These classical string equations clearly show that
the classical scattering amplitudes,
in both the pictures,  are insensitive to
deformations of the parameters.

\section{ Nonrelativistic Fermion}

          The matrix model becomes, after diagonalizing and double-scaling,
a theory of free nonrelativistic fermions in an inverted oscillator potential
deformed by $1/x^2$.
The hamiltonian of second-quantized fermions is given by
\begin{eqnarray}
H=
\int_0^{+\infty}dx~
      \Psi^*(x)
           \left\{ -\frac{1}{2}\partial_x^2
                   +V(x)-\mu
            \right\} \Psi(x),
\label{1}
\end{eqnarray}
where the potential $V(x)$ is
\begin{eqnarray}
V(x)=-\frac{1}{2}x^2+\frac{m}{2x^2}~~,~~~~~~~~m > 0~~,
\label{2}
\end{eqnarray}
and $\mu$ is the Fermi energy.
$m$ is a deformation parameter of this nonrelativistic fermion system.
$\Psi(x)$ and $\Psi^*(x)$ are second-quantized fermion fields
with $x$  the rescaled matrix eigenvalue.
Their equal time anti-commutation
relations are set to
\begin{eqnarray}
\{~ \Psi^*(x_1)~,~ \Psi(x_2) ~\} &=& \delta(x_1-x_2)~~, \nonumber \\
\{ ~\Psi^*(x_1)~, ~\Psi^*(x_2) ~\} &=&
          \{ ~\Psi(x_1)~, ~\Psi(x_2) ~\}~=~0~~ .
\label{3}
\end{eqnarray}
Notice that the integration in the R.H.S of
equation (\ref{1}) is over the region
$\mbox{$[$}0, +\infty )$.
This is due to the  appearance of an
infinite wall of the potential
$V(x)$ at the origin. Then  wave functions of the one-body system will
be defined on this
region and vanish at the origin $x=0$, which can be considered as
a boundary condition for
second-quantized system (\ref{1}).


\subsection{Algebraic properties}

        We shall begin by studying  algebraic properties  associated with
hamiltonian
(\ref{1}),
which will play an important role in the description of the
underlying Euclidean
$c=1$ strings.
For this purpose we shall look into the one-body  hamiltonian operator
$L_x=-\frac{1}{2}\partial_x^2+V(x)$.
In this quantum mechanical system
it may be convenient
to introduce
the following second-order differential operators
\begin{eqnarray}
B_x&=&\frac{1}{2}(\partial_x-ix)^2-\frac{m}{2x^2}~~, \nonumber \\
\bar{B}_x&=&\frac{1}{2}(\partial_x+ix)^2-\frac{m}{2x^2}~~.
\label{4}
\end{eqnarray}
These two operators give an  analogue of
spectral generating algebra for this quantum
mechanical system.
They satisfy the commutation relations
\begin{eqnarray}
\mbox{$[$}~L_x~,~ B_x ~\mbox{$]$}&=& 2i B_x ~~,\nonumber \\
\mbox{$[$}~L_x~,~ \bar{B}_x ~\mbox{$]$}&=& -2i \bar{B}_x~~, \nonumber \\
\mbox{$[$}~B_x~,~ \bar{B}_x ~\mbox{$]$}&=& -4i L_x~~.
\label{5}
\end{eqnarray}
So one can construct the series of hamiltonian eigenstates by
the successive operations of
$B_x (\bar{B}_x)$ to the ground-state wave function.
Notice that with an appropriate rescaling of the operators these
commutation relations
are reduced to those of $su(1,1)$ algebra :
$\mbox{$[$}h,f \mbox{$]$}
=-if,\mbox{$[$}h,e \mbox{$]$} =ie,
\mbox{$[$}e,f \mbox{$]$}=-2ih$.
Since the Casimir element of $su(1,1)$ is given by
$\Omega=h^2-(ef+fe)/2$,
the corresponding operator $\Omega_x$ becomes constant
\begin{eqnarray}
\Omega_x &\equiv& \frac{1}{4}\left(
              L_x^2-\frac{B_x\bar{B}_x+\bar{B}_xB_x}{2} \right) ~~,
\nonumber \\
        &=& \frac{1}{4}\left(\frac{3}{4}-m \right)~~.
\label{7}
\end{eqnarray}
To summarize one may say that  algebraic nature
of this quantum mechanical
system is governed by the envelops of $su(1,1)$ algebra
(with Casimir condition (\ref{7})).

               Let us return to  nonrelativistic fermion system (\ref{1}).
It should be remarked first
that an algebra of differential operators such as (\ref{5}) lifts up
to the second-quantized
form. In fact, for differential operators
$D_x^{(i)} (i=1,2)$, we can construct  second-quantized operators
$\hat{D}^{(i)}$ as
\begin{eqnarray}
\hat{D}^{(i)}~=~\int dx~ \Psi^*(x)D_x^{(i)} \Psi(x)~~.
\end{eqnarray}
The commutation relation among these operators satisfies
\begin{eqnarray}
\mbox{$[$}~\hat{D}^{(1)}~,~\hat{D}^{(2)}~\mbox{$]$}
{}~=~\widehat{ \mbox{$[$} D^{(1)},D^{(2)} \mbox{$]$} }~~,
\label{8}
\end{eqnarray}
which tells us that the original algebra
is preserved  in the second-quantized form.
We shall introduce the second-quantized operators which correspond to
$B_x$ and $\bar{B}_x$
\begin{eqnarray}
B_1 &=& \int_0^{\infty}dx~\Psi^*(x)B_x \Psi(x), \nonumber \\
\bar{B}_1 &=& \int_0^{\infty}dx~ \Psi^*(x)\bar{B}_x \Psi(x).
\label{9}
\end{eqnarray}
$su(1,1)$ algebra (\ref{5}) lifts up in the second-quantized  form
\begin{eqnarray}
\mbox{$[$}~H~,~ B_1 ~\mbox{$]$}&=& 2i B_1 ~~,\nonumber \\
\mbox{$[$}~H~,~ \bar{B}_1 ~\mbox{$]$}&=& -2i \bar{B}_1~~, \nonumber \\
\mbox{$[$}~B_1~,~ \bar{B}_1~ \mbox{$]$}&=& -4i(H +\mu N)~~,
\label{10}
\end{eqnarray}
where $N=\int_0^{\infty}dx \Psi^*\Psi(x)$ is the fermion number operator.
Other elements of the $su(1,1)$ enveloping algebra in the one-body system
may be described by the following set of  second-quantized operators
\begin{eqnarray}
O_{J,M}^a
{}~\equiv~\int_0^{\infty}dx~ \Psi^*(x)
   L_x^{~a}\bar{B}_x^{~\frac{J+M}{2}}B_x^{~\frac{J-M}{2}} \Psi(x)~~,
\label{11}
\end{eqnarray}
where $a, J \in Z_{\geq 0}$ and $M=-J,-J+2,\cdots,J-2,J$.
%Clearly these operators
%form the $su(1,1)$ enveloping algebra.
(Because of  Casimir relation (\ref{7}) all
these operators are not independent.)
One can also say these operators
as the fermionic realization of the $W_{\infty}$ algebra
\cite{Winfty} which appears in the collective field analysis
\cite{JY,AJ} of the deformed
matrix model.
In this sense,
as we shall verify it in the later section, the operators
\begin{eqnarray}
B_n~\equiv ~O^{a=0}_{n,-n}~,~~~~~\bar{B}_n~\equiv ~O^{a=0}_{n,n}~~
\label{12}
\end{eqnarray}
will be identified with the creation
operators of the massless tachyon states with momentum $\pm 2n$
in the Euclidean $c=1$ strings.

%%%%%%%%%%%%%%%%%%%%%%%%%%%%%%%%%%%%%%%%%%%%%%%%%%%%%%%%%%%%%%%%%%%%%%%%%%%

\subsection{Analytic properties}

          Let us consider  analytical properties  of  nonrelativistic
fermion system (\ref{1}).

         Firstly we shall look into the effect of
the boundary condition imposed by the infinite wall
of the potential $V(x)$.
The eigenvalue problem for the one-body hamiltonian operator
$L_x$ $(\equiv -\frac{1}{2}\partial_x^2+V(x))$
\begin{eqnarray}
L_x u_{\xi}(x)~=~ \xi u_{\xi}(x)
\label{14}
\end{eqnarray}
will be solved by using the Whittaker function.
Since the Whittaker function $W(z)( \equiv M_{\kappa,\pm \mu}(z),
W_{\pm \kappa,\mu}(\pm z) )$
\footnote{
We follow the normalization in \cite{GR}
\begin{eqnarray*}
M_{\kappa,\mu}(x)
&=&
x^{\mu +\frac{1}{2}}e^{-\frac{x}{2}}F(\mu -\kappa +\frac{1}{2},2 \mu +1;x)~~,
\nonumber \\
W_{\kappa,\mu}(x) &=&
\frac{\Gamma(-2 \mu)}{\Gamma(\frac{1}{2}- \mu -\kappa)}M_{\kappa,\mu}(x)
+\frac{\Gamma(2 \mu)}{\Gamma(\frac{1}{2}+ \mu -\kappa)}M_{\kappa,-\mu}(x)~~,
\end{eqnarray*}
where $F(a,b;x)$ is a degenerate hypergeometric function.
}
satisfies the differential equation
\begin{eqnarray}
\left\{ \partial^2_z+
         \left( -\frac{1}{4}+
               \frac{\kappa}{z}-\frac{\mu^2-\frac{1}{4}}{z^2}\right)
\right\} W(z)~=~0~~~,
\end{eqnarray}
the following can be taken as the independent solutions for this eigenvalue
problem (\ref{14})
\begin{eqnarray}
x^{-\frac{1}{2}}M_{\frac{\xi}{2i},\pm \alpha}(i x^2)~~~~~
\mbox{or}~~~~~~~x^{-\frac{1}{2}}W_{\pm \frac{\xi}{2i}, \alpha}(\pm i x^2)~~~ ,
\label{15}
\end{eqnarray}
where we set
\begin{eqnarray}
\alpha=\frac{1}{4}\sqrt{1+4m}.
\label{16}
\end{eqnarray}
The behaviors at the origin $x=0$ of these eigenfunctions are
\begin{eqnarray}
x^{-\frac{1}{2}}M_{\frac{\xi}{2i},\pm \alpha}(i x^2)&=&
e^{\frac{\pi i }{2}(\frac{1}{2}\pm \alpha)}e^{-\frac{i}{2}x^2}x^{\pm 2 \alpha}
\mbox{$[$}1 + O(x^2) \mbox{$]$}~~, \nonumber \\
x^{-\frac{1}{2}}W_{\pm \frac{\xi}{2i}, \alpha}(\pm i x^2)&=&
e^{\mp \frac{i}{2}x^2}
   \left\{
     \frac{ \Gamma(-2\alpha)e^{\pm \frac{\pi i}{2}(\frac{1}{2}+\alpha)}}
            {\Gamma(\frac{1}{2}-\alpha \mp \frac{\xi}{2i})}
        x^{2 \alpha}
      +
     \frac{ \Gamma(2\alpha)e^{\pm \frac{\pi i}{2}(\frac{1}{2}-\alpha)}}
            {\Gamma(\frac{1}{2}+\alpha \mp \frac{\xi}{2i})}
        x^{-2 \alpha}
    \right\}
         \mbox{$[$}1 + O(x^2) \mbox{$]$}~~. \nonumber \\
&& ~~~~~~
\label{17}
\end{eqnarray}
Since $m>0$ implies $\alpha=\sqrt{1+4m}/4>1/4$, the physical
boundary condition, $u_{\xi}(x)|_{x=0}=0$, will be satisfied only for
\begin{eqnarray}
x^{-\frac{1}{2}}M_{\frac{\xi}{2i},\alpha}(i x^2)~~.
\label{19}
\end{eqnarray}
Thus it is very conceivable that any wave function can
be expanded by the eigenfunctions
$x^{-\frac{1}{2}}M_{\frac{\xi}{2i},\alpha}(i x^2)$.
Say,
for an example, we may expand the fermion fields $\Psi(x), \Psi^*(x)$ as
\begin{eqnarray}
\Psi(x)&=&
\int d \xi~ \Psi_{-\xi}
     x^{-\frac{1}{2}}M_{\frac{\xi}{2i},\alpha}(i x^2)~~,~~
\nonumber \\
\Psi^*(x)&=&
\int d \xi~ \Psi^*_{-\xi}
       x^{-\frac{1}{2}}M_{\frac{\xi}{2i},\alpha}(i x^2)~~.
\label{35}
\end{eqnarray}

           Notice that,
from the linear independence of solutions (\ref{15}),
the wave function
$x^{-\frac{1}{2}}M_{\frac{\xi}{2i},\alpha}(i x^2)$
can be also written as
the linear combination of
$x^{-\frac{1}{2}}W_{\pm \frac{\xi}{2i}, \alpha}(\pm i x^2)$
\begin{eqnarray}
&&x^{-\frac{1}{2}}M_{\frac{\xi}{2i},\alpha}(i x^2) \nonumber \\
&& =
\Gamma(2\alpha+1)e^{-\frac{\pi \xi}{2}}
\left\{
    \frac{e^{\frac{\pi i}{2}(1+2 \alpha)}}
              {\Gamma(\alpha+\frac{1}{2}+\frac{\xi}{2i})}
        x^{-\frac{1}{2}}W_{\frac{\xi}{2i},\alpha}(ix^2)
   +\frac{1}{\Gamma(\alpha+\frac{1}{2}-\frac{\xi}{2i})}
         x^{-\frac{1}{2}}W_{-\frac{\xi}{2i},\alpha}(-ix^2)
\right\}.\nonumber \\
&&~~~~~~
\label{20}
\end{eqnarray}
The asymptotic behavior of the wave function
$x^{-\frac{1}{2}}M_{\frac{\xi}{2i},\alpha}(i x^2)$ can be read from those of
$x^{-\frac{1}{2}}W_{\pm \frac{\xi}{2i}, \alpha}(\pm i x^2)$ \cite{GR}
\begin{eqnarray}
x^{-\frac{1}{2}}W_{\pm \frac{\xi}{2i}, \alpha}(\pm i x^2)
\sim
e^{\frac{\pi \xi}{4}}x^{-\frac{1}{2}}
e^{\mp \frac{i}{2}x^2 \mp i \xi \ln x}
\mbox{$[$}1+O(\frac{1}{x^2}) \mbox{$]$}~~,
\label{18}
\end{eqnarray}
which is valid for $-\frac{\pi}{4}< Arg x< \frac{\pi}{4}$.

          Nextly we shall consider  eigenvalue problem (\ref{14})
from  rather different points of view.
For this purpose we will
introduce unitary transforms on the space of the one-body
wave functions. These transforms will be used in the construction of
the asymptotic fields
for $\Psi(x)$ and $\Psi^*(x)$.
The underlying classical mechanics of the hamiltonian operator $L_x$ is
described by
\begin{eqnarray}
\frac{\partial x}{\partial \tau}=p~,~~~
\frac{\partial p}{\partial \tau}=
-\frac{\partial V(x)}{\partial x}~~.
\label{21}
\end{eqnarray}
Especially on the dynamical orbit with the energy
$E(\equiv \frac{p^2}{2}+V(x))$ equal to the Fermi energy $\mu$,
the measure $d \tau$ has the form
\begin{eqnarray}
d \tau =
\frac{dx}{\sqrt{2(\mu-V(x))}}~~,
\label{22}
\end{eqnarray}
from which one can see that the R.H.S of (\ref{22}) is the
1-form invariant under the shift $\tau \rightarrow \tau+\delta \tau$
on this orbit.
The integration of (\ref{22})
\begin{eqnarray}
\tau =
\int_{x_0}^{x}
\frac{dy}{\sqrt{2(\mu-V(y)}}~~,~~~~~~~~(x \geq x_0),
\label{23}
\end{eqnarray}
where $x=x_0$ is the turning point of the dynamical orbit $E=\mu$,
maps the region
$x_0 \leq x < +\infty$ onto the region $0 \leq \tau <+\infty$
bijectively.
Let us introduce similarity transforms
$\cal{J}_{\pm}$ which map the wave functions $u(x)$
\footnote{
Strictly speaking we should restrict the class of the wave functions to those
which have their support on $x > x_0$.}
to the functions on
the region $0 \leq \tau < +\infty$
\begin{eqnarray}
\mbox{$\cal{J}$}_{\pm}~~:~~u(x) \mapsto \mbox{$\cal{J}$}_{\pm}u(\tau)
\equiv J_{\pm}(x(\tau))u(x(\tau))~~~,
\label{24}
\end{eqnarray}
where
\begin{eqnarray}
J_{\pm}(x)=
  \{ 2(\mu-V(x)) \}^{\frac{1}{4}}
      e^{\pm i \int_{x_0}^{x}\sqrt{2(\mu-V(y))}dy}~~.
\end{eqnarray}
Notice that, especially for the wave functions
which support are on the region $x > x_0$,
these similarity transforms $\cal{J}_{\pm}$ are unitary,
that is, preserve the  $L^2$-norm,
\begin{eqnarray}
|\mbox{$\cal{J}$}_{\pm}u|^2_{\tau}
&\equiv& \int_0^{+\infty}d \tau ~~
\overline{\mbox{$\cal{J}$}_{\pm}u(\tau)}
\mbox{$\cal{J}$}_{\pm}u(\tau) ~~,\nonumber \\
&=&
\int_0^{+\infty}d \tau ~~\frac{dx}{d\tau}
 \overline{u(x(\tau))}u(x(\tau))~~, \nonumber \\
&=&
\int_0^{+\infty}dx~~ \overline{u(x)}u(x)~~, \nonumber \\
&\equiv&
|u|^2_x~~~~~.
\end{eqnarray}

       Under these  transforms  the one-body  hamiltonian  operator $L_x$
will change
into $\tilde{L}_{\tau}^{\pm}$. Their explicit forms are given by
\begin{eqnarray}
 \tilde{L}_{\tau}^{\pm}&\equiv& J_{\pm}L_xJ_{\pm}^{-1}~~, \nonumber \\
&=&
\pm i \partial_{\tau}+\mu
  -\frac{1}{2}f^{-\frac{1}{2}}
\partial_{\tau}f^{-1}\partial_{\tau}f^{-\frac{1}{2}}~~~,
\label{25}
\end{eqnarray}
where $f(\tau)=\sqrt{2(\mu-V(x(\tau)))}$.
One can also consider the eigenvalue
problems for these transformed operators
$\tilde{L}_{\tau}^{\pm}$,
\begin{eqnarray}
\tilde{L}_{\tau}^{\pm}v_{\xi}^{\pm}(\tau)
      ~ =~\xi v_{\xi}^{\pm}(\tau)~~.
\label{26}
\end{eqnarray}
Though these eigenvalue problems are eqivalent
to (\ref{14}) via the correspondence
$v_{\xi}^{\pm}(\tau)=\mbox{$\cal{J}$}_{\pm}u_{\xi}(\tau)$,
we may consider them
perturbatively. Namely we divide $\tilde{L}_{\tau}^{\pm}$ into
\begin{eqnarray}
\tilde{L}_{\tau}^{\pm}&=&
(\pm i \partial_{\tau}+\mu ) +\tilde{L}_{\tau}^{int}~~, \nonumber \\
\left( \tilde{L}_{\tau}^{int} \right. &\equiv& \left.
   -\frac{1}{2}f^{-\frac{1}{2}}\partial_{\tau}f^{-1}
\partial_{\tau}f^{-\frac{1}{2}}
\right)
\label{27}
\end{eqnarray}
and then treat the second order differential operator
$\tilde{L}_{\tau}^{int}$
as a perturbation to $\pm i \partial_{\tau}+\mu $.
The zero-th order approximation for  eigenvalue problems (\ref{26})
gets the form
\begin{eqnarray}
(\pm i \partial_{\tau}+\mu )v_{\xi}^{\pm~(0)}(\tau)
{}~=~ \xi v_{\xi}^{\pm~(0)}(\tau)~~,
\end{eqnarray}
from which one can see
\begin{eqnarray}
v_{\xi}^{\pm~(0)}(\tau)~=~e^{\mp i (\xi-\mu)\tau}~~.
\label{28}
\end{eqnarray}
Under the inverse transforms $\mbox{$\cal{J}$}_{\pm}^{-1}$
the zero-th order
solutions $v_{\xi}^{\pm~(0)}(\tau)$
are mapped to
\begin{eqnarray}
\mbox{$\cal{J}$}_{\pm}^{-1}v_{\xi}^{\pm~(0)}(x)=
\frac{1}{\{ 2(\mu-V(x)) \}^{\frac{1}{4}}}
e^{\mp i \int_{x_0}^{x}\sqrt{2(\mu-V(y))}dy}
e^{\mp i (\xi-\mu)\tau(x)},
\label{29}
\end{eqnarray}
which asymptotic behaviors can be read as
\footnote{The following estimates are useful
\begin{eqnarray*}
&&
\int_{x_0}^{x}\frac{dy}{\sqrt{2(\mu-V(y))}}
{}~\sim~
\ln x -\frac{1}{4}\ln \frac{\mu^2+m}{4}+ O(\frac{1}{x^2}) ~~ ,\\
&&
\int_{x_0}^x \sqrt{2(\mu-V(y))}dy
{}~\sim~
\frac{x^2}{2}+\mu \ln x-\frac{\mu}{4}\ln
\frac{\mu^2+m}{4} +a+ O(\frac{1}{x^2})~~ .
\end{eqnarray*}
where
$a \equiv
\frac{\mu}{2}-\frac{\sqrt{m}}{2}
    (\frac{\pi}{2}+Arcsin \frac{\mu}{\sqrt{\mu^2+m}})$.
}
\begin{eqnarray}
\mbox{$\cal{J}$}_{\pm}^{-1}v_{\xi}^{\pm~(0)}(x)
{}~\sim~
e^{\mp ia}e^{\pm \frac{i\xi}{4}\ln \frac{\mu^2+m}{4}}
x^{-\frac{1}{2}}e^{\mp i \frac{x^2}{2}\mp i \xi \ln x}
\mbox{$[$}1+ O(\frac{1}{x^2}) \mbox{$]$}~~,
\end{eqnarray}
where $a$ is the constant independent of $\xi$ (and is irrelevant to our
discussion). These  asymptotic behaviors coincide
(up to the leading order) with those
of $x^{-\frac{1}{2}}W_{\pm \frac{\xi}{2i},\alpha}(\pm i x^2)$ (\ref{18})
by  constant multiplications.
So one may expect that
the eigenfunctions $v_{\xi}^{\pm}(\tau)$
which are constructed from $v_{\xi}^{\pm (0)}(\tau)$ by the
perturbative expansion, that is,
$v_{\xi}^{\pm}(\tau)$
$\equiv v_{\xi}^{\pm (0)}(\tau)
+v_{\xi}^{\pm (1)}(\tau)+ \cdots $, lead to
the asymptotic expansion of
$x^{-\frac{1}{2}}W_{\pm \frac{\xi}{2i},\alpha}(\pm i x^2)$
\begin{eqnarray}
e^{\mp ia-\frac{\pi \xi}{4}\pm \frac{i \xi}{4}\ln \frac{\mu^2+m}{4}}
 x^{-\frac{1}{2}}W_{\pm \frac{\xi}{2i},\alpha}(\pm i x^2)
{}~\sim~ \mbox{$\cal{J}$}_{\pm}^{-1}v_{\xi}^{\pm}(x)~~.
\label{32}
\end{eqnarray}

                     We shall proceed to the
the second-quantized system.
            Under  similarity transforms
$\mbox{$\cal{J}$}_{\pm}$ (\ref{24})
second-quantized hamiltonian operator $H$ (\ref{1})
will become
\begin{eqnarray}
\tilde{H}_{\pm}&=&
\tilde{H}_{\pm}^{(0)}+\tilde{H}_{\pm}^{int}~~, \nonumber \\
\tilde{H}_{\pm}^{(0)}&=&
\int_{0}^{+\infty}d \tau ~
\tilde{\Psi}^*_{\pm}(\tau)
(\pm i \partial_{\tau})
\tilde{\Psi}_{\pm}(\tau) ~~,\nonumber \\
\tilde{H}_{\pm}^{int}&=&
\int_{0}^{+\infty}d \tau ~
\tilde{\Psi}^*_{\pm}(\tau)
\tilde{L}_{\tau}^{int}
\tilde{\Psi}_{\pm}(\tau)~~,
\label{33}
\end{eqnarray}
where
$\tilde{H}_{\pm}^{(0)}$ can be considerd as
the analogue of the hamiltonian for
$1+1$ dimensional free relativistic fermions \cite{SW,GK}.

               With the assumption  that the contributions  of
$\tilde{H}_{\pm}^{int}$ at
$\tau \gg 0$ are negligible
it may be possible to introduce the asymptotic fields
$\tilde{\Psi}_{in,out}(\tau)$ by
\begin{eqnarray}
\tilde{\Psi}_{in}(\tau)&=&
\int d \xi ~\psi_{-\xi}^{in}v_{\xi}^{+ (0)}(\tau)~~ ,\nonumber \\
\tilde{\Psi}_{out}(\tau)&=&
\int d \xi ~\psi_{-\xi}^{out}v_{\xi}^{- (0)}(\tau)~~,
\label{34}
\end{eqnarray}
such that
$\tilde{\Psi}_+(\tau)\rightarrow \tilde{\Psi}_{in}(\tau),$
$\tilde{\Psi}_-(\tau)\rightarrow \tilde{\Psi}_{out}(\tau) $
as $\tau \rightarrow +\infty$.
It is important to note that these asymptotic fields
$\tilde{\Psi}_{in,out}(\tau)$
are not independent to each other.
Due to  expansion (\ref{35})
of $\Psi(x)$ by the
Whittaker functions,
each mode $\psi_{-\xi}^{in,out}$ in (\ref{34}) is related  with the mode
$\Psi_{-\xi}$  in (\ref{35}) through
asymptotic relation (\ref{32}) and  decomposition (\ref{20})
of the one-body eigenfunction.
This constraint gives us the following relation between
$\psi_{-\xi}^{in}$ and $\psi_{-\xi}^{ out}$.
\begin{eqnarray}
\psi_{-\xi}^{in}=
e^{i(2a+\frac{\pi}{2}(1+2\alpha))}
e^{-i \frac{\xi}{2} \ln \frac{\mu^2+m}{4}}
\frac{\Gamma(\frac{1}{2}+\alpha+\frac{i}{2}\xi)}
        {\Gamma(\frac{1}{2}+\alpha-\frac{i}{2}\xi) }
\psi^{out}_{-\xi}~~.
\label{36}
\end{eqnarray}
We can also repeat the silmilar discussion
for the conjugate field $\Psi^*(x)$.
By introducing the asymptotic fields as
\begin{eqnarray}
\tilde{\Psi}^*_{in}(\tau)&=&
\int d \xi~ \psi_{\xi}^{* in}\overline{v_{\xi}^{+ (0)}(\tau)} ~~,\nonumber \\
\tilde{\Psi}^*_{out}(\tau)&=&
\int d \xi ~\psi_{\xi}^{* out}\overline{v_{\xi}^{- (0)}(\tau)}~~,
\end{eqnarray}
where the bar  denotes the complex conjugation,
the relation between the modes $\psi_{\xi}^{* in,out}$ can be read as
\begin{eqnarray}
\psi_{\xi}^{* in}=
e^{-i(2a+\frac{\pi}{2}(1+2\alpha))}
e^{+i \frac{\xi}{2} \ln \frac{\mu^2+m}{4}}
\frac{\Gamma(\frac{1}{2}+\alpha-\frac{i}{2}\xi)}
        {\Gamma(\frac{1}{2}+\alpha+\frac{i}{2}\xi) }
\psi^{out}_{\xi}~~.
\label{37}
\end{eqnarray}
Notice that, from the forms of these asymptotic fields ,
the (non-vanishing) anti-commutation relations
among these in-coming (out-going) fields
can be expressed in terms of their modes.
\begin{eqnarray}
\left\{~ \psi_{\xi_1}^{in}~,~\psi_{\xi_2}^{* in}~ \right\}
&=&\delta(\xi_1+\xi_2) ~~
\nonumber \\
\left\{~ \psi_{\xi_1}^{out}~,~ \psi_{\xi_2}^{* out}~ \right\}
&= &\delta(\xi_1+\xi_2)~~.
\end{eqnarray}
Scattering relations (\ref{36}) and (\ref{37}) are of course consistent
with these anti-commutation relations.

%%%%%%%%%%%%%%%%%%%%%%%%%%%%%%%%%%%%%%%%%%%%%%%%%%%%%%%%%%%%%%%%%%%%%%

\subsection{Euclidean  strings}

             The scattering amplitudes of the Euclidean strings
may be described by the analytic continuation of the theory
to the Euclidean region. This prescription for the analytic continuation
was explained in \cite{MPR}. The energy $\xi$ of the fermion modes
$\psi_{-\xi}^{in,out}$ and $\psi_{\xi}^{* in, out}$ will
replace $\xi-\mu \rightarrow iq$, where $q \in R$.
{}From now on we shall write these fermion modes as
$\psi_{-q}^{in,out}$ and $\psi_{q}^{* in, out}$.
After this continuation,  relations (\ref{36}) and (\ref{37})
between the in-coming and out-going fermion modes may be written as
\begin{eqnarray}
\psi_q^{* in} &=& r(q)\psi_q^{* out}~~,
\nonumber \\
\psi_{-q}^{in} &=& r(q)^{-1}\psi_{-q}^{ out} ~~,
\label{39}
\end{eqnarray}
where we introduce $r(q)$,
\begin{eqnarray}
r(q) ~=~
\left(
\frac{4}{\mu^2+m}
\right)^{\frac{q}{2}}
\frac{\Gamma(\frac{1}{2}+\alpha-i \frac{\mu}{2}+\frac{q}{2})}
         {\Gamma(\frac{1}{2}+\alpha+i \frac{\mu}{2}-\frac{q}{2})}~~.
\end{eqnarray}
In this expression
we absorb the $q-$independent phases which appear
in (\ref{36}) and (\ref{37})
into the normalizations of the asymptotic fields
$\tilde{\Psi}_{in,out}(\tau)$ and $\tilde{\Psi}^{*}_{in,out}(\tau)$ .
The value of $r(q)$ coincides with that obtained in
\cite{DK}
from the
asymptotic behavior of the resolvent kernel ,
$<x| \frac{1}{L+\mu+iq} |y>$, and,
as is expected  from the form of the one-body hamiltonian
$L_x$,
taking $m \rightarrow 0$, the value of $r(q)$ becomes
equal to that for the type I theory
of the inverse harmonic oscilator potential,
$-x^2/2$,  with an infinite wall at the origin
\cite{MPR}.

    Let us introduce the in-coming and out-going vacuums
of these asymptotic fermions  by the condition
\begin{eqnarray}
< in |\psi_q^{in}=0, ~~< in |\psi_q^{* in}=0
      &&~~~ \mbox{for}~~q<0~~, \nonumber \\
\psi_q^{out}|out>=0, ~~\psi_q^{* out}|out>=0
      &&~~~ \mbox{for}~~q>0~~.
\end{eqnarray}
The scattering amplitudes among the nonrelativistic fermions
will be reduced to the
combination of the
following matrix elements
\begin{eqnarray}
<in|
\prod_{i}\psi^{in}_{p_i}
\prod_{j}\psi^{* in}_{q_j}
\prod_{k}\psi^{* out}_{r_k}
\prod_{l}\psi^{out}_{s_l} |out>~~.
\label{nonmat}
\end{eqnarray}
It is important to note that these matrix elements can be also realized
in terms of two-dimensional
relativistic fermion \cite{MPR}. In fact  matrix element (\ref{nonmat})
is equal to
\begin{eqnarray}
<0|
\prod_{i}\psi_{p_i}
\prod_{j}\psi^{*}_{q_j}
g
\prod_{k}\psi^{*}_{r_k}
\prod_{l}\psi_{s_l} |0>~~,
\end{eqnarray}
where
$\psi_q$ and $\psi^*_q$  are the relativistic fermions
with their (non-vanishing) anti-commutation relation,
$~\left\{ \psi_{q_1}, \psi_{q_2}^{*} \right\}= \delta(q_1+q_2)~$, and
the ground-state is introduced by the condition,
\begin{eqnarray*}
<0|\psi_q=0, ~~<0|\psi_q^{*}=0 &&~~~ \mbox{for}~~q<0~~, \nonumber \\
\psi_q|0>=0, ~~\psi_q^{*}|0>=0 &&~~~ \mbox{for}~~q>0~~.
\end{eqnarray*}
$~g~$, an element of $GL(\infty)$, is introduced such that
it transforms the relativistic
fermions as
\begin{eqnarray}
g^{-1}~\psi_q^{*}~g &=& r(q) ~\psi^*_q~~~~,~~~\nonumber \\
g^{-1}~\psi_{-q} ~g &=& r(q)^{-1}~ \psi_{-q} ~~~~,
\label{40}
\end{eqnarray}
for $\forall q$.
Notice that  relation (\ref{39}) is now translated to the adjoint action
of $~g~$ on the relativistic fermions  $\psi_q$
and $\psi^*_q$.

         Nextly let us consider the asymptotic forms of  operators
$B_n$ and $\bar{B}_n$ (\ref{12}).
{}From the definitions of these operators (\ref{11})
their actions on the nonrelativistic fermions are
\begin{eqnarray}
\mbox{$[$}~ B_n~,~ \Psi(x) ~\mbox{$]$} &=& - B_x^{~n}\Psi(x) ~~,\nonumber \\
\mbox{$[$} ~\bar{B}_n~,~ \Psi(x) ~\mbox{$]$} &=& - \bar{B}_x^{~n}\Psi(x)~~ .
\label{44}
\end{eqnarray}
The actions of the $2n$-th differential operators
$B_x^{~n}
\left( \equiv \left(
\frac{1}{2}(\partial_x-ix)^2-\frac{m}{2x^2} \right)^n \right)$
and its complex conjugate
$\bar{B}_x^{~n}$
%$\left( \equiv \left(
%\frac{1}{2}(\partial_x+ix)^2-\frac{m}{2x^2} \right)^n \right) $
on the nonrelativistic fermions
(the R.H.Ss of (\ref{44}))
can be analyzed further from their actions on
one-body wave function
$x^{- \frac{1}{2}}M_{\frac{\xi}{2i},\alpha}(ix^2)$
(\ref{19}).
Taking account of the Euclidean continuation we may expand the
nonrelativistic fermion field
$\Psi(x)$ by the Whittaker function
$x^{- \frac{1}{2}}M_{-\frac{i \mu}{2}+\frac{q}{2},\alpha}(ix^2)$
\begin{eqnarray}
\Psi(x)=
\int dq~ \Psi_{-q}
x^{- \frac{1}{2}}M_{-\frac{i \mu}{2}+\frac{q}{2},\alpha}(ix^2)~~.
\label{41}
\end{eqnarray}
The actions of $B_x$ and $\bar{B}_x$ on the wave function
$x^{- \frac{1}{2}}
         M_{-\frac{i \mu}{2}+\frac{q}{2},\alpha}(ix^2)$ can be obtained
from the recursion relations
\footnote{
The recursion relations are \cite{GR}
\begin{eqnarray*}
x \partial_xM_{\kappa,\mu}(x)
{}~=~
\pm(\frac{x}{2}-\kappa)M_{\kappa,\mu}(x)+
(\mu \pm \kappa + \frac{1}{2})M_{\kappa \pm 1,\mu}(x).
\end{eqnarray*}
$su(1,1)$ structure (\ref{5}) of the one-body system
has its origin in these recurrent relations of the Whittaker function.
}
of the Whittaker function
\begin{eqnarray}
B_xx^{- \frac{1}{2}}M_{-\frac{i \mu}{2}+\frac{q}{2},\alpha}(ix^2)&=&
-2i \left(
\frac{1}{2}+\alpha -\frac{i \mu}{2} +\frac{q}{2}
\right)
x^{- \frac{1}{2}}
       M_{-\frac{i \mu}{2}+\frac{q+2}{2},\alpha}(ix^2)~~~ , \nonumber \\
\bar{B}_xx^{- \frac{1}{2}}M_{-\frac{i \mu}{2}+\frac{q}{2},\alpha}(ix^2)&=&
2i
\left(
\frac{1}{2}+\alpha +\frac{i \mu}{2} -\frac{q}{2}
\right)
x^{- \frac{1}{2}}M_{-\frac{i \mu}{2}+\frac{q-2}{2},\alpha}(ix^2)~~~ .
\label{411}
\end{eqnarray}
Then we can evaluate the both sides of
equations (\ref{44}) so that we can see from (\ref{44}) that  $B_n$ and
$\bar{B}_n$ transform the fermion modes $\Psi_{-q}$ as
\begin{eqnarray}
\mbox{$[$}~ B_n~,~\Psi_{-q}~ \mbox{$]$}
&=&
-(-2i)^{n}
\frac{\Gamma(\frac{1}{2}+\alpha-\frac{i \mu}{2}+\frac{q}{2})}
        {\Gamma(\frac{1}{2}+\alpha-\frac{i \mu}{2}+\frac{q-2n}{2})}
           \Psi_{-q+2n} ~~,
\nonumber \\
\mbox{$[$}~ \bar{B}_n~, \Psi_{-q} ~\mbox{$]$}
&=&
-(2i)^{n}
\frac{\Gamma(\frac{1}{2}+\alpha+\frac{i \mu}{2}-\frac{q}{2})}
        {\Gamma(\frac{1}{2}+\alpha+\frac{i \mu}{2}-\frac{q+2n}{2})}
           \Psi_{-q-2n}~~.
\label{45}
\end{eqnarray}
These actions on $\Psi_{-q}$ will be
handed over on the asymptotic modes $\psi_{-q}^{in. out}$.
Using  decomposition (\ref{20}),
fermionic field $\Psi(x)$
(\ref{41}) can be written as the  sum
\footnote{We introduce the following quantites
\begin{eqnarray*}
s_+(q)&=&
       e^{\frac{\pi i}{2}( i\mu -q+1+2\alpha)}
             \frac{\Gamma(1+2\alpha)}
                     {\Gamma(\frac{1}{2}+\alpha-\frac{i\mu}{2}+\frac{q}{2})}
  \left( \frac{4}{\mu^2+m} \right)^{-\frac{q}{4}} ~~, \\
s_-(q)&=&
        e^{\frac{\pi i}{2}( i\mu -q)}
         \frac{\Gamma(1+2\alpha)}
              {\Gamma(\frac{1}{2}+\alpha+\frac{i\mu}{2}-\frac{q}{2})}
        \left( \frac{4}{\mu^2+m} \right)^{\frac{q}{4}}~~~~ .
\end{eqnarray*}
}
\begin{eqnarray}
\Psi(x)&=&
\int dq ~
s_+(q) \Psi_{-q}
{}~\left( \frac{4}{\mu^2+m} \right)^{\frac{q}{4}}x^{-\frac{1}{2}}
W_{-\frac{i \mu}{2}+\frac{q}{2}, \alpha} (ix^2)
\nonumber \\
&& +
\int dq ~
s_-(q) \Psi_{-q}
{}~\left( \frac{4}{\mu^2+m} \right)^{-\frac{q}{4}}x^{-\frac{1}{2}}
W_{\frac{i \mu}{2}-\frac{q}{2}, \alpha} (-ix^2)~~~ .
\label{42}
\end{eqnarray}
Each integration in the R.H.S of (\ref{42}) will reduce to  the asymptotic
fields
$~(\mbox{$\cal{J}$}_{\pm})^{-1}
\tilde{\Psi}_{in , out }(x)~$
$=~\int dq ~ \psi_{-q}^{in , out}
(\mbox{$\cal{J}$}_{\pm})^{-1}v_{\mu+iq}^{\pm (0)}(x)~$.
In fact ,
by using the explicit correspondence between the wave functions (\ref{32})
\begin{eqnarray}
\left( \frac{4}{\mu^2+m} \right)^{\pm \frac{q}{4}}x^{-\frac{1}{2}}
W_{\mp\frac{i \mu}{2}\pm \frac{q}{2}, \alpha} (\pm ix^2)
{}~\sim~
(\mbox{$\cal{J}$}_{\pm})^{-1}v_{\mu+iq}^{\pm}(x)~~,
\end{eqnarray}
one can derive the  relations between these several fermion modes
\begin{eqnarray}
\psi_{-q}^{in}
&=&
\frac{e^{-\frac{i \pi q}{2}}}
     {\Gamma(\frac{1}{2}+\alpha-\frac{i \mu}{2}+ \frac{q}{2})}
\left( \frac{4}{\mu^2+m} \right)^{-\frac{q}{4}}
\Psi_{-q} ~~,
\nonumber \\
\psi_{-q}^{out}
&=&
\frac{e^{-\frac{i \pi q}{2}}}
     {\Gamma(\frac{1}{2}+\alpha+\frac{i \mu}{2}- \frac{q}{2})}
\left( \frac{4}{\mu^2+m} \right)^{\frac{q}{4}}
\Psi_{-q}~~ .
\label{43}
\end{eqnarray}
The asymptotic actions of $B_n$ and $\bar{B}_n$ will be obtained
from  transforms (\ref{45}) by changing $\Psi_q$ to
$\psi_q^{in,out}$ through  correspondence (\ref{43})
\begin{eqnarray}
\mbox{$[$}~ B_n^{in}~, ~\psi_{-q}^{in}~\mbox{$]$}
&=&
-(2i)^n \left( \frac{\mu^2+m}{4}\right)^{\frac{n}{2}}
\psi_{-q+2n}^{in} ~~,\nonumber \\
\mbox{$[$}~ \bar{B}_n^{in}~,~ \psi_{-q}^{in} ~\mbox{$]$}
&=&
-(-2i)^n \left( \frac{\mu^2+m}{4}\right)^{\frac{n}{2}}
\frac{r(q+2n)}{r(q)}
\psi_{-q-2n}^{in}~~ , \nonumber \\
\mbox{$[$}~ B_n^{out}~,~ \psi_{-q}^{out}~ \mbox{$]$}
&=&
-(2i)^n \left( \frac{\mu^2+m}{4}\right)^{\frac{n}{2}}
\frac{r(q)}{r(q-2n)}
\psi_{-q+2n}^{out}~~ ,\nonumber \\
\mbox{$[$}~ \bar{B}_n^{out}~,~ \psi_{-q}^{out} ~\mbox{$]$}
&=&
-(-2i)^n \left( \frac{\mu^2+m}{4}\right)^{\frac{n}{2}}
\psi_{-q-2n}^{out} ~~~~.
\label{46}
\end{eqnarray}
Since $B_n$ and $\bar{B}_n$ have fermion bilinear forms (\ref{11})
these actions completely determine their forms.
It may be also convenient to write them
in terms of the relativisitc fermions. Let us introduce
the operators
\begin{eqnarray}
\mbox{$\cal{B}$}_n &\equiv&
-(2i)^n
\left( \frac{\mu^2+m}{4} \right)^{\frac{n}{2}}
\int dp~ \psi_{-p}\psi^*_{p+2n} ~~,
\nonumber \\
\bar{\mbox{$\cal{B}$}}_n &\equiv&
-(-2i)^n
\left( \frac{\mu^2+m}{4} \right)^{\frac{n}{2}}
\int dp~ \psi_{-p}\psi^*_{p-2n} ~~,
\label{47}
\end{eqnarray}
for $n >0$.
Then we may
write $B_n^{in, out}$ and $\bar{B}_n^{in, out}$ as
\begin{eqnarray}
B^{in}_n~=~\mbox{$\cal{B}$}_n~, ~~
&&~~\bar{B}_n^{in}~=~g \bar{\mbox{$\cal{B}$}}_n g^{-1}~~,
\nonumber \\
B^{out}_n~=~g^{-1}\mbox{$\cal{B}$}_n g~, ~~
&&~~\bar{B}_n^{out}=\bar{\mbox{$\cal{B}$}}_n~~ ,
\label{48}
\end{eqnarray}
which tell us the scattering relations of
$B_n$ and $\bar{B}_n$.
{}From expressions (\ref{47})
we can see that
$\mbox{$\cal{B}$}_n $ and $\bar{\mbox{$\cal{B}$}}_n $ are
essentially the (even) modes
of a relativistic free boson field
$\partial \phi$ or equivalently
$B_n^{in}$ and $\bar{B}_{n}^{out}$ are the (even) modes
of asymptotic (free) boson fields
$\partial \phi_{in. out}$. In the next section we will show
that these asymptotic boson fields can be exactly identified
with those
appeared in the collective field approach.


\section{Compactification at  Self-Dual Radius}

                                     In this section we shall study
the Euclidean strings  compactified at self-dual radius.
The prescription for the compactification was explained in \cite{KL,DMP}.
At the compactification radius $\beta$
the Euclidean momentum $q \in R$ of the relativistic fermions
$\psi_{q}, \psi^*_{q}$ are discretized to
$\frac{n+\frac{1}{2}}{\beta}~~(n \in Z)$ .
In particular, at the self-dual radius $\beta=1$,
action  (\ref{40}) of $g$ on the relativistic
fermions gets the form
\begin{eqnarray}
g^{-1}\psi^*_n g &=& e^{\alpha_n}\psi^*_n ~~, \nonumber \\
g^{-1}\psi_{-n} g &=& e^{-\alpha_n}\psi_{-n}~~  ,
\label{51}
\end{eqnarray}
where we set
\begin{eqnarray}
e^{\alpha_n}& \equiv &r(n+\frac{1}{2}) ~~, \nonumber \\
&=&
 \left( \frac{4}{\mu^2+m} \right)^{\frac{n+\frac{1}{2}}{2}}
 \frac{\Gamma(\frac{3}{4}+\alpha-i\frac{\mu}{2}+\frac{n}{2})}
        {\Gamma(\frac{1}{4}+\alpha+i\frac{\mu}{2}-\frac{n}{2})} ~~.
\end{eqnarray}
Notice that, for the notational convenience,
we shift the label of the fermion modes by $1/2$.
We shall summarize our usage of the relativistic fermions :
$\psi(z)=\sum_n\psi_n z^{-n}$ and $\psi^*(z)=\sum_n\psi^*_nz^{-n-1}$
are relativistic fermion fields
\footnote{
Strictly speaking, under this shift of modes,
$(\psi,\psi^*)$  is the  ghost system
with central charge $c=-2$.}
with the nontrivial operator
product expansion,
$\psi(z)\psi^*(w)\sim \frac{1}{z-w}$,
or equivalently
$\{\psi_n,\psi^*_m\}=\delta_{n+m,0}$.
Their ground-state $|n>$ $(n \in Z)$ is introduced by the condition
\begin{eqnarray}
\psi_m |n> &=& 0~~~~~~m>-n~~, \nonumber \\
\psi^*_m|n> &=& 0~~~~~~m \geq n ~~.
\end{eqnarray}
$<n|$ is dual to $|n>$, that is, $<n|m>=\delta_{n,m}$, and it
satisfies the condition
\begin{eqnarray}
<n|\psi_m &=& 0~~~~~~m\leq -n~~,\nonumber \\
<n|\psi^*_m &=& 0~~~~~~m<n~~.
\end{eqnarray}

%%%%%%%%%%%%%%%%%%%%%%%%%%%%%%%%%%%%%%%%%%%%%%%%%%%%%%%%%%%%%%%%%%%%%
%%%%%%%%%%%%%%%%%%%%%%%%%%%%%%%%%%%%%%%%%%%%%%%%%%%%%%%%%%%%%%%%%%%%%

\subsection{$su(1,1)$ structure}

      Let us begin by studying the implications of
relations (\ref{48}) between
$\bar{B}_n^{in}$ and $\bar{B}_n^{ out}$
(or $B_n^{ out}$ and $B_n^{ out}$ ) in these compactified strings.
For this purpose it may be convenient to review the free fermion
realization of  pseudo-differential operators
\cite{bib:DJKM}.
We shall introduce
a infinite dimensional vector space $\cal{V}$ with basis
$\{e_n~ (n \in Z) \}$,
on which $X$ and $\partial_X$ act as
$Xe_n=(n+1)e_{n+1}$ and $\partial_X e_n=e_{n-1}$.
$\psi_n$ and $\psi^*_{n}$ have the following
realization  on $\wedge \mbox{$\cal{V}$}$,
\begin{eqnarray}
\psi_n &=& e_{-n} \wedge~~, \nonumber \\
\psi^*_n &=& i_{e_n} ~~~~,
\end{eqnarray}
where  $v \wedge$ and $i_v$ denote respectively
the exterior and interior products by $v \in $ $\cal{V}$.
With this representation of the fermions
any polynomial  of $X$, $\partial_X$ and
$\partial_X^{-1}$ gets the fermion bilinear form.
\footnote{ The action of pseudo-differential operators on
$\cal{V}$
lifts up on $\wedge \mbox{$\cal{V}$}$ with a standard
normal ordering prescription. }
Especially  pseudo-differential operators,  $X^k\partial_X^{~n+k}$,
have the form
\begin{eqnarray}
:X^k\partial_X^{~n+k}: ~~=~~
\sum_{p=- \infty}^{+ \infty} k!
\left( \begin{array}{c}p \\ k \end{array} \right)
:\psi_{-p}\psi^*_{p+n}: ~~~~~.
\label{52}
\end{eqnarray}
The normal ordering $~:~~:~$ is prescribed relative to the ground-state
$|0>\in$ $\wedge$ $\cal{V}$, and
$\left( \begin{array}{c}p \\ k \end{array} \right)
\equiv \frac{p(p-1)\cdots(p-k+1)}{k!}$
is the binomial coefficient.
These (normal ordered) actions of
pseudo-differential operators on  $\wedge$$\cal{V}$ constitute
$W_{1+\infty}$ algebra
\footnote{With our realization
$~\{ W_n^{(k)}\}~$
include the Virasoro algebra
$~\{  W_n^{(1)} \}~$ with central charge equal to
$-2$.},
a subalgebra of $\hat{gl}(\infty)$ algebra.
Notice that the R.H.S of (\ref{52}) is a mode of the higher spin
field $W^{(k)}(z) \equiv : \partial^{~k}_z \psi \psi^* :(z)$.
By introducing the mode expansion for this higher spin field as
\begin{eqnarray}
W^{(k)}(z)~=~
\sum_{n=- \infty}^{+\infty} ~
W_n^{(k)} z^{-n-k-1}~~,
\label{53}
\end{eqnarray}
the R.H.S of (\ref{52}) becomes equal to $W_n^{(k)}$.
Therefore the generators of this $W_{1+\infty}$ algebra
%\footnote{With our realization
%$~\{ W_n^{(k)}\}~$
%include the Virasoro algebra
%$~\{  W_n^{(1)} \}~$ with central charge equal to
%$-2$.}
can be also written as
\begin{eqnarray}
W_n^{(k)}~=~:X^k\partial_X^{~n+k}:~~~.
\label{54}
\end{eqnarray}

            With this brief review we shall return to our problem.
We notice that $\mbox{$\cal{B}$}_n$ and $\bar{\mbox{$\cal{B}$}}_n$
(\ref{47})
become essentially equal to
$W_{2n}^{(0)}$ and $W_{-2n}^{(0)}$ after this compactification.
Then, from  realizations (\ref{48}) of $B_n^{in,out}$ and
$\bar{B}_n^{in,out}$,
one can see that
the relations between $\bar{B}_n^{in}$ and $\bar{B}_n^{ out}~$
(or$~B_n^{in}$ and $B_n^{ out}~$) are
reduced to those between
$W_{-2n}^{(0)}$ and $g W_{-2n}^{(0)} g^{-1}~$
(or$~W_{2n}^{(0)}$ and $g^{-1} W_{2n}^{(0)} g~$).
So it may be interesting
to describe $g W_{-2n}^{(0)} g^{-1}~$ (or$~g^{-1} W_{2n}^{(0)} g~$)
in terms of the generators
of $W_{1+\infty}$ algebra.
In particular, for the case of $n=1$,
$g W_{-2}^{(0)} g^{-1} $ gets
the form
\begin{eqnarray}
 g ~W_{-2}^{(0)} ~g^{-1} &=&
\sum_p  :g \psi_{-p}g^{-1} \cdot g \psi^*_{p-2} g^{-1}:~~, \nonumber \\
&=&
\sum_p \frac{e^{\alpha_p}}{e^{\alpha_{p-2}}} :\psi_{-p}\psi^*_{p-2}:~~,
\nonumber \\
&=&
\frac{1}{\mu^2+m}
\left\{
(m-i \mu (i\mu+1))W_{-2}^{(0)}
+2i \mu W_{-2}^{(1)}
-W_{-2}^{(2)} \right\} ~~.
\label{55}
\end{eqnarray}
Through  correspondence (\ref{54}) one can also interpret this equation
by pseudo-differential operators
\begin{eqnarray}
g ~:\partial_X^{~-2}:~ g^{-1} ~=~
\frac{1}{\mu^2+m}
: ~\left\{
-(X-i\mu \partial_X^{~-1})^{2}
+m \partial_X^{~-2} \right\}~ :~~~.
\label{56}
\end{eqnarray}
As for $g^{-1} W_{2}^{(0)} g$,
it has the form
\begin{eqnarray}
g^{-1}~ W_{2}^{(0)}~ g &=&
\frac{1}{\mu^2+m}
\left\{
(m-(i \mu -1)(1\mu -2))W_{2}^{(0)}
+2(i \mu-2) W_{2}^{(1)}
-W_{2}^{(2)} \right\},
\label{74}
\end{eqnarray}
or equivalently
\begin{eqnarray}
g^{-1}~:\partial_X^{~2}:~ g ~=~
\frac{1}{\mu^2+m}
: ~\left\{
-(X \partial_{X}^{~2}+(1-i\mu) \partial_X)^{2}
+m \partial_X^{~2} \right\}~ :~~~.
\label{75}
\end{eqnarray}
Notice that, as we will clarify it in the next section ,
these two equations
(\ref{56}) and (\ref{75})  constitute
string equations for the deformed $c=1$ strings.

                  Explicit forms  (\ref{55}) (or (\ref{74})) of
$gW_{-2}^{(0)}g^{-1}$  (or $g^{-1}W_2^{(0)}g$)
are useful to investigate the $su(1,1)$ structure of  compactified
$c=1$ strings.
Let us introduce the operator $\cal{H}$ as
\begin{eqnarray}
\mbox{$\cal{H}$}~\equiv~
\frac{-i}{\sqrt{\mu^2+m}}
\left\{
W_0^{(1)}
+\left( \frac{1}{2}-i \mu \right) W_0^{(0)}
-\frac{m+\mu^2}{2}
\right\}~~~.
\label{851}
\end{eqnarray}
Then $W_2^{(0)},gW_{-2}^{(0)}g^{-1}$ and $\cal{H}$  forms
the $su(1,1)$ algebra
\begin{eqnarray}
\mbox{$[$}~W_2^{(0)} ~,~gW_{-2}^{(0)}g^{-1} ~\mbox{$]$}&=&
\frac{-4i}{\sqrt{\mu^2+m}}~~\mbox{$\cal{H}$}~~~~,\nonumber \\
\mbox{$[$}~\mbox{$\cal{H}$} ~,~W_2^{(0)} ~\mbox{$]$}&=&
\frac{2i}{\sqrt{\mu^2+m}}~ ~W_2^{(0)} ~~~~,\nonumber \\
\mbox{$[$}~\mbox{$\cal{H}$} ~,~gW_{-2}^{(0)}g^{-1} ~\mbox{$]$}&=&
\frac{-2i}{\sqrt{\mu^2+m}} ~~gW_{-2}^{(0)}g^{-1} ~~~~.
\label{85}
\end{eqnarray}
The above commutators are  evaluated, after substituting
the R.H.S of (\ref{55}) for $gW_{-2}^{(0)}g^{-1}$,
by utilizing the following commutation relations of $W_{1+\infty}$
algebra \cite{bib:Sato-Sato}
\begin{eqnarray}
&&  \mbox{$[$}~W_m^{(k)} ~,~W_n^{(l)} ~\mbox{$]$} \nonumber \\
&& = \sum_{r \geq 0} \left\{
\left( \begin{array}{c}k+m \\ r \end{array} \right)
\left( \begin{array}{c}k+l-r \\ k \end{array} \right)-
\left( \begin{array}{c}l+n \\ r \end{array} \right)
\left( \begin{array}{c}k+l-r\\ l \end{array} \right)
\right\}
\frac{k!l!}{(k+l-r)!}
W_{m+n}^{(k+l-r)} \nonumber \\
&&~~~~~~~+(-1)^kk!l! \left( \begin{array}{c}k+m \\ k+l+1 \end{array} \right)
\delta_{m+n,0}~~~~~~~,
\end{eqnarray}
which are obtainable from the operator product expansions
among the  higher
spin fields $W^{(k)}(z)$ realized by the free fermion.
Since $\cal{H}$ (\ref{851}) is invariant under the adjoint action of
$g$, we can equivalently
construct the $su(1,1)$ algebra by
$g^{-1}W_2^{(0)}g$, $W_{-2}^{(0)}$ and
$\cal{H}$
\begin{eqnarray}
\mbox{$[$}~g^{-1}W_2^{(0)}g ~,~W_{-2}^{(0)} ~\mbox{$]$}&=&
\frac{-4i}{\sqrt{\mu^2+m}}~~\mbox{$\cal{H}$}~~~~,\nonumber \\
\mbox{$[$}~\mbox{$\cal{H}$} ~,~g^{-1}W_2^{(0)}g ~\mbox{$]$}&=&
\frac{2i}{\sqrt{\mu^2+m}} ~~g^{-1}W_2^{(0)}g ~~~~,\nonumber \\
\mbox{$[$}~\mbox{$\cal{H}$} ~,~W_{-2}^{(0)} ~\mbox{$]$}&=&
\frac{-2i}{\sqrt{\mu^2+m}} ~~W_{-2}^{(0)} ~~~~.
\label{84}
\end{eqnarray}

                     This appearance of  $su(1,1)$  algebra can be
considered  as the result of  same algebraic structure (\ref{5})
revealed  in the study
of the one-body hamiltonian operator $L_x$.
More precisely, since  $su(1,1)$ structure (\ref{5}) has its origin
in the recursion relations of the Whittaker function
and the element $g \in GL(\infty)$ itself is essentially determined
from the properties
of this function, one can say that
$su(1,1)$ structure (\ref{85}) in these  compactified
$c=1$ strings
comes from the Whittaker
function.

           For the case of $n \geq 2$, the explicit description of
$gW_{-2n}^{(0)}g^{-1} \left( \equiv g:\partial_X^{-2n}:g^{-1} \right)$
and $g^{-1}W_{2n}^{(0)}g \left( \equiv g^{-1}:\partial_X^{2n}:g \right)$
by the generators of $W_{1+\infty}$ algebra is very complicated.
It may be instructive to give their forms in terms of pseudo-differential
operators
\begin{eqnarray}
g ~:\partial_X^{~-2n}:~ g^{-1} &=&
\left(\frac{1}{\mu^2+m}\right)^n
: ~\left\{
-(X-i\mu \partial_X^{~-1})^{2}
+m \partial_X^{~-2} \right\}^n~ :~~~, \nonumber \\
g^{-1}~:\partial_X^{~2n}:~ g &=&
\left(\frac{1}{\mu^2+m}\right)^n
: ~\left\{
-(X \partial_{X}^{~2}+(1-i\mu) \partial_X)^{2}
+m \partial_X^{~2} \right\}^n~ :~~.
\label{742}
\end{eqnarray}
The R.H.Ss of these equations can be translated to the sums of
$W^{(l)}_{\mp 2n}$ according to  (\ref{54}).
Taking account of these expressions of
$gW_{-2n}^{(0)}g^{-1}$ (or $g^{-1}W_{2n}^{(0)}g$)
besides commutation relations (\ref{85}) ( or (\ref{84})),
one may see that these elements $gW_{-2n}^{(0)}g^{-1}$
and $W_{2m}$
(or $g^{-1}W_{2n}^{(0)}g$ and $W_{-2m}$) with their combinations
are giving a representation of (central extended) enveloping
algebra of $su(1,1)$ .
Therefore one can
construct
a representation of $W_{\infty}$ algebra from
these elements.

        In order to proceed to the next subsection
it may be convenient to
give some remarks related with this $W_{\infty}$ symmetry.
Because
the operators $:X^k\partial_X^{n+k}:$ and
$g^{-1}:X^k\partial_X^{n+k}:g$
will constitute $W_{1+\infty}$ algebras respectively,
let us prepare two $W_{1+\infty}$ algebras.
For the definitive we shall denote them as
$W_{1+\infty}^{(+)}$  and
$W_{1+\infty}^{(-)}$,
each of which is generated by $:X^k\partial_X^{n+k}:$
or $:Y^k\partial_Y^{n+k}:$.
These two $W_{1+\infty}$ algebras are not independent.
They are related to each other by the adjoint action of $g$
\begin{eqnarray}
W_{1+\infty}^{(-)}&=&g^{-1}W_{1+\infty}^{(+)}g~~~,\nonumber \\
\left(^{i.e}~ W_n^{(k)~(-)} \right.
&=&
\left.g^{-1}W_n^{(k)~(+)}g ~\right)~~~.
\label{quantumscattering}
\end{eqnarray}
$su(1,1)$ algebra (\ref{85}) (or (\ref{84}))
can be regarded as the consequence
of
intertwining
(\ref{quantumscattering})
between these two $W_{1+\infty}$ algebras.
Therefore
our $W_{\infty}$ algebra
is also the result of this intertwining.
Notice that
$W_{1+\infty}$ algebra
can be realized  in terms of free bosons, that is,
the bosonization of the higher spin fields
$:\partial_z^k \psi \psi^*:(z)$ $\left( \equiv W^{(k)}(z) \right)$.
By introducing a free boson field
$\partial_z \phi(z)$ with the operator product expansion
$\partial_z\phi(z)\partial_w\phi(w) \sim \frac{1}{(z-w)^2}$,
the higher spin field gets the form
\begin{eqnarray}
W^{(k)}(z)=
   \frac{1}{k+1}:\left\{
    \left( \partial_z\phi \right)^{k+1}+\cdots \right\}:(z)~~~,
\label{bosonrep}
\end{eqnarray}
where "$\cdots$" denotes the quantum corrections. Especially,
in the cases of $k=0,1$ and $2$, $W^{(k)}(z)$ have the following
bosonized forms
\begin{eqnarray}
W^{(0)}(z) &=& \partial \phi(z) ~~~, \nonumber \\
W^{(1)}(z) &=&
\frac{1}{2}:\left( (\partial \phi)^2+\partial^2\phi \right):(z)~~~,
\nonumber \\
W^{(2)}(z) &=&
\frac{1}{3}:\left( (\partial \phi)^3+3\partial \phi \partial^2\phi
+\partial^3\phi \right):(z)~~~.
\end{eqnarray}
These two $W_{1+\infty}$ algebras,
$W_{1+\infty}^{(\pm)}$, will be realized by  boson fields
$\partial \phi_{+}$ and $\partial \phi_-$ respectively.
It is important to note that,
under these bosonizations of
$W_{1+\infty}^{(\pm)}$,
intertwining relation
(\ref{quantumscattering}) will replace the scattering relations
between $\partial \phi_{\pm}$. As we will see it in the next subsection,
these scattering relations are preceisely the nonperturbative
analogue of Polchinski's classical scattering equations.

%%%%%%%%%%%%%%%%%%%%%%%%%%%%%%%%%%%%%%%%%%%%%%%%%%%%%%%%%%%%%%%%%%%%%%%%
%%%%%%%%%%%%%%%%%%%%%%%%%%%%%%%%%%%%%%%%%%%%%%%%%%%%%%%%%%%%%%%%%%%%%%%%

\subsection{Classical limits and Polchinski's scattering equations}

               In  this subsection we will
consider the properties of  "classical limits" of
the several equations
obtained in the last subsection.
Classical limit may be taken by the following substitution
under $\hbar \rightarrow 0$
\begin{eqnarray}
\hbar \partial_X &\rightarrow& P~~~~, \nonumber \\
\mbox{$[$}~
\hbar \partial_X~,~X~
\mbox{$]$}~=~ \hbar
&\rightarrow&
\{~P~,~Q ~\}_{p.b.}~=~1~~,
\end{eqnarray}
where we identify $X$ with $Q$.
The Poisson bracket
is given by
$\{ F, G \}_{p.b.}$
$=~\frac{\partial F}{\partial P}\frac{\partial G}{\partial Q}$
$-\frac{\partial G}{\partial P}\frac{\partial F}{\partial Q}.$
The problem about "taking classical limits" of the equations
in the previous subsection lies on the
ambiguity in identifying the Planck constant
$\hbar$ with the parameters which appear in these equations.

        Let us first consider the classical limit under the  identification
\begin{eqnarray}
\hbar ~\equiv~
\frac{-1}{i \sqrt{m}}~~  .
\label{60}
\end{eqnarray}
With this identification
one may rewrite  equations (\ref{56}) and (\ref{75}) as
\begin{eqnarray}
(1-\mu^2\hbar^2)g:\left(\hbar \partial_X \right)^{-2}:g^{-1}
&=&
:\left\{ \left(X-i\mu\hbar \left(\hbar \partial_x \right)^{-1}\right)^{2}
+\left( \hbar \partial_X \right)^{-2} \right\}: ~~~~,
\nonumber \\
(1-\mu^2\hbar^2)g^{-1}:\left(\hbar \partial_X \right)^{2}:g
&=&
: \left\{
\left(X \left(\hbar \partial_X \right)^2+(1-i \mu)\hbar
\left(\hbar \partial_X \right) \right)^2+
\left(\hbar \partial_X \right)^2 \right\}:~.
\label{88}
\end{eqnarray}
Then the $\hbar = -1/(i \sqrt{m})  \rightarrow 0$ limits of these equations
become independent of $\mu$
\begin{eqnarray}
P_+^{~-2}&=&Q^{~2}_-+P^{~-2}_- ~~,\nonumber \\
P_-^{~2} &=& Q_+^{~2}P_+^{~4}+P_+^{~2}~~,
\label{61}
\end{eqnarray}
where we introduce $P_{\pm}$ and $Q_{\pm}$ as the classical counterparts
of the
following operators,
\begin{eqnarray}
P_+~ \sim~ : \hbar \partial_X :~,&&~~~Q_+~ \sim~ : X :~~,\nonumber \\
P_-~ \sim~ g^{-1}: \hbar \partial_X :g~,&&~~~Q_- ~\sim~ g^{-1} : X : g~~~.
\end{eqnarray}
Notice that these classical limits (\ref{61})
preserve the Poisson bracket
$\{P,Q\}_{p.b.}=1$. This is the classical analogue
of the simple fact that the adojoint actions of $g$
on $X$ and $\hbar \partial_X$ preserve their
commutator.
%$\mbox{$[$} g^{-1}\hbar \partial_X g,g^{-1} X g \mbox{$]$}=
% \mbox{$[$} \hbar \partial_X , X \mbox{$]$} (=\hbar)$.
So they define a symplectic diffeomorphism
on  two-plane, that is, on the classical
phase space
$(Q,P)$.
On the other hand these quantities besides
$Q_-P_-$ (or equivalently $Q_+P_+$) form the
$su(1,1)$ Poisson algebra
\begin{eqnarray}
\left\{~P_-^{~2}~,~P_+^{~-2}~\right\}_{p.b.}
&=& 4~Q_-P_-~~~, \nonumber \\
\left\{~Q_-P_-~,~P_-^{~2}~\right\}_{p.b.}
&=& -2~P_-^{~2}~~~,  \nonumber \\
\left\{~Q_-P_-~,~P_+^{~-2}~\right\}_{p.b.}
&=& 2~P_+^{~-2}~~~,
\label{86}
\end{eqnarray}
which is the classical limit of
(\ref{85}) ((\ref{84})).
Equations (\ref{742}) will replace under this limit
\begin{eqnarray}
P_+^{~-2n}~=~\left( Q^{~2}_-+P^{~-2}_- \right)^n ~,~
P_-^{~2n} ~=~ \left( Q_+^{~2}P_+^{~4}+P_+^{~2} \right)^n~~,
\label{62}
\end{eqnarray}
where the former is the limit of
$:\partial_X^{-2n}:$ described in terms of
$g^{-1}:X^l \partial_X^{l-2n}:g$ and the latter
is that of $g^{-1}:\partial_X^{2n}: g$
evaluated by $:X^l \partial_X^{l+2n}:$.
So it follows that, under the Poisson bracket,
polynomials of
$Q_-P_-,P_+^{~-2}$ and $P_-^{~2}$
constitute the enveloping algebra of $su(1,1)$,
that is,  $w_{\infty}$ algebra.
It is the classical limit of the $W_{\infty}$ algebra
discussed in the previous subsection.

          The another identification will be
\begin{eqnarray}
\hbar ~=~ \frac{-1}{i \mu}~~~.
\label{57}
\end{eqnarray}
With this identification  equations
(\ref{56}) and (\ref{75}) get the forms
\begin{eqnarray}
(1-m \hbar^2)g:\left( \hbar \partial_X \right)^{-2}:g^{-1}
&=&
: \left\{ \left(X + \left( \hbar \partial_X \right)^{-1} \right)^2
-m \hbar^2 \left( \hbar \partial_X \right)^{-2} \right\}:~~~~,\nonumber \\
(1-m \hbar^2)g^{-1}:\left(\hbar \partial_X \right)^{2}:g
&=&
: \left\{
\left(X \left(\hbar \partial_X \right)^2+(1+\hbar)
\left(\hbar \partial_X \right) \right)^2-m \hbar^2
\left(\hbar \partial_X \right)^2 \right\}:,
\label{89}
\end{eqnarray}
from which we can read the
$\hbar  = -1/(i \mu) \rightarrow 0$ limits of these equations as
\begin{eqnarray}
P_+^{~-2} &=& (Q_-+P_-^{~-1})^2~~,\nonumber \\
P_-^{~2} &=& (Q_+P_+^{~2}+P_+)^2~~,
\label{58}
\end{eqnarray}
which are independent of $m$.
Notice that equations (\ref{58}) define another symplectic
diffeomorphism on the classical phase spce.
It also follows that
$P_+^{-2},P_-^{~2}$ and $Q_-P_-+1$ form the $su(1,1)$ Poisson
algebra
\begin{eqnarray}
\{~P_-^{~2}~,~P_+^{~-2}~\}_{p.b.}&=& 4(Q_-P_-+1) ~~, \nonumber \\
\{~Q_-P_-+1~,~P_-^{~2}~\}_{p.b.} &=& -2P_-^{~2}~~, \nonumber \\
\{~Q_-P_-+1~,~P_+^{~-2}~\}_{p.b.} &=& 2P_+^{~-2}~~.
\label{87}
\end{eqnarray}
This Poisson algebra is also the classical analogue
of (\ref{85}).
Since the classical limits of  equations (\ref{742}) are
\begin{eqnarray}
P_+^{-2n} ~=~ (Q_-+P^{-1}_-)^{2n}~~, ~~~~
P_-^{2n} ~=~ (Q_+P_+^2+P_+)^{2n}~~~~,
\label{59}
\end{eqnarray}
polynomials of $P_+^{-2},P_-^2$  and $Q_-P_-$ constitute
$w_{\infty}$ algebra under the Poisson bracket.
It is also the classical limit of the $W_{\infty}$ algebra
obtained in the last subsection.

             At this stage it may be convenient to give some remarks
on these identifications of $\hbar$
from the string theoretical point of view.
In string theory the genus expansion
of the free energy may be considered as
the asymptotic expansion by $\hbar$.
The leading contribution is that of the classical string.
So the above classical limits
will be related with
the classical (genus zero) contributions of string .
The different identifications of $\hbar$
will lead to the different asymptotic expansions of the nonperturbative
free energy, hence,
the different string models.
Our first identification (\ref{60}) is based
on the proposal by Jevicki and Yoneya \cite{JY}, in which they tried
to search the nonperturbative nature of string in the black-hole background.
On the other hand our second choice is the standard identification in
the $c=1$ string \cite{DMP}.
As we have shown, these exists the $w_{\infty}$ symmetry in both classical
limits. But their constituents are different from each other.

                  Since  equations (\ref{56}) and
(\ref{75}) are equivalent to  expressions (\ref{55}) and
(\ref{74}),
it may be also interesting to
study these equations from the view of
the classical limit of  $W_{1+\infty}$ algebra.
The classical limit
of $W_n^{(k)}$ will be given by
\begin{eqnarray}
w_n^{(k)}~\equiv ~Q^kP^{n+k}~~.
\label{63}
\end{eqnarray}
They constitute the following Poisson algebra
\begin{eqnarray}
\{~ w_n^{(k)}~,~ w_m^{(l)} ~\}_{p.b.}
{}~=~(ln-km)w_{n+m}^{(k+l-1)}~~.
\label{64}
\end{eqnarray}
This algebra is preceisely the algebra of  area-preserving
(or symplectic) diffeomorphisms
on two-plane, that is, $w_{1+\infty}$ algebra.
As $W_{1+\infty}$ algebra has the realization by free bosons,
its classical limit will be also written by
using a classical boson field.
Let us introduce a classical boson field
$\alpha(y)=\sum_{n=-\infty}^{+\infty} \alpha_n e^{-iny}$
with the Poisson structure,
$~\{ \alpha(y_1),\alpha(y_2) \}$
$=2 \pi i \partial_y \delta(y_1-y_2)$ .
Generators  (\ref{63}) of $w_{1+\infty}$ algebra are given by
\begin{eqnarray}
w_n^{(k)}~=~
\frac{1}{k+1}\int \frac{dy}{2 \pi i}
{}~~e^{iny} \alpha(y)^{k+1}~~ ,
\label{66}
\end{eqnarray}
which are the classical counterpart  of the bosonic realization of
$W_n^{(k)}$.

         We shall write down our two different classical equations
(\ref{62}) and (\ref{59})
in terms of these classical bosons. Since we have two $w_{1+\infty}$ algebras
constructed respectively from  the pairs $(P_+,Q_+)$ and $(P_-,Q_-)$,
that is, the classical limits of $W^{(\pm)}_{1+\infty}$ which
are described at the end of the last subsection,
it may be convenient to introduce
two classical boson fields for their realizations
\begin{eqnarray}
\alpha_{\pm}(y)~ = ~\sum_{n=-\infty}^{+\infty} \alpha_n^{(\pm)}e^{-iny}~~.
\end{eqnarray}
Let us first consider
classical equations (\ref{62})
which are derived from  (\ref{75}) and (\ref{56})
under
$\hbar  = -1/(i \sqrt{m}) \rightarrow 0$ limit.
Using $\alpha_{\pm}(y)$, the R.H.Ss of these equations (\ref{62})
can be evaluated into
\begin{eqnarray}
&& (Q^{~2}_-+P^{~-2}_-)^n \nonumber \\
&&=
\sum_{r \geq 1}\frac{\Gamma(1+n)}{\Gamma(2+n-r)(r-1)!}
Q_-^{~2r-2}P_-^{~-2n+2r-2}~~,
\nonumber \\
&&=
\sum_{r \geq 1}\frac{\Gamma(1+n)}{(2r-1)\Gamma(2+n-r)(r-1)!}
\int \frac{dy}{2 \pi i} e^{-2iny}\alpha_-(y)^{2r-1}~~, \nonumber \\
&&=
\sum_{r \geq 1}\frac{\Gamma(1+n)}{(2r-1)\Gamma(2+n-r)(r-1)!}
\sum_{m_1,\cdots,m_{2r-1}}
\delta_{2n+\sum_{j=1}^{2r-1}m_j, 0}
\alpha_{m_1}^{(-)}\cdots \alpha_{m_{2r-1}}^{(-)}~~, \nonumber \\
&&~~~~~~
\end{eqnarray}
and
\begin{eqnarray}
&&(Q^{~2}_+P_+^{~4}+P^{~2}_+)^n  \nonumber \\
&&=
\sum_{r \geq 1}\frac{\Gamma(1+n)}{\Gamma(2+n-r)(r-1)!}
Q_+^{~2r-2}P_+^{~2n+2r-2} ~~,
\nonumber \\
&&=
\sum_{r \geq 1}\frac{\Gamma(1+n)}{(2r-1)\Gamma(2+n-r)(r-1)!}
\int \frac{dy}{2 \pi i} e^{2iny}\alpha_+(y)^{2r-1} ~~,\nonumber \\
&&=
\sum_{r \geq 1}\frac{\Gamma(1+n)}{(2r-1)\Gamma(2+n-r)(r-1)!}
\sum_{m_1,\cdots,m_{2r-1}}
\delta_{-2n+\sum_{j=1}^{2r-1}m_j, 0}
\alpha_{m_1}^{(+)}\cdots \alpha_{m_{2r-1}}^{(+)}~~. \nonumber \\
&&~~~~~~~~~~
\end{eqnarray}
Thus we obtain the  bosonic realizations for these classical
equations (\ref{62})
\begin{eqnarray}
\alpha_{-2n}^{(+)} &=&
\sum_{r \geq 1}\frac{\Gamma(1+n)}{(2r-1)\Gamma(2+n-r)(r-1)!}
\sum_{m_1,\cdots,m_{2r-1}}
\delta_{2n+\sum_{j=1}^{2r-1}m_j, 0}
\alpha_{m_1}^{(-)}
\cdots \alpha_{m_{2r-1}}^{(-)} ~~,
\nonumber \\
\alpha_{2n}^{(-)}
&=&
\sum_{r \geq 1}\frac{\Gamma(1+n)}{(2r-1)\Gamma(2+n-r)(r-1)!}
\sum_{m_1,\cdots,m_{2r-1}}
\delta_{-2n+\sum_{j=1}^{2r-1}m_j, 0}
\alpha_{m_1}^{(+)}\cdots \alpha_{m_{2r-1}}^{(+)} . \nonumber \\
&&~~~
\label{69}
\end{eqnarray}

          Though the above equations have been derived in the framework of
the compactified
string, it may be very plausible that the similar equations  also hold
for the noncompact case
by simply changing the sum of momentums in the R.H.Ss  of (\ref{69}) to
the integrations
\begin{eqnarray}
\alpha_{q}^{(\pm)} ~=~
\sum_{r \geq 1}
\frac{\Gamma(1 \mp \frac{q}{2})}
                   {(2r-1)\Gamma(2 \mp \frac{q}{2}-r)(r-1)!}
\int \prod_{j=1}^{2r-1} dp_j
\delta(-q+\sum_{j=1}^{2r-1}p_l)
\alpha_{p_1}^{(\mp)}\cdots \alpha_{p_{2r-1}}^{(\mp)}~~.
\label{691}
\end{eqnarray}
It is important to note that,
by analytically continuing to the Minkowski region,
these equations  (\ref{691}) are giving  formal solutions of
Polchinski's scattering
equations
\begin{eqnarray}
\alpha_{\pm}(y)~=~\alpha_{\mp}(y \mp \frac{1}{2}
\ln (1+ \alpha_{\pm}^2(y))) ~~.
\label{71}
\end{eqnarray}
These scattering equations were derived  in \cite{JY}
from  the asymptotic behavior of the classical
collective fields $\tilde{\alpha}_{\pm}(t,x)$ which satisfy the equation of
motion
\begin{eqnarray}
\partial_t \tilde{\alpha}_{\pm}~=~
-\partial_x V(x)-\tilde{\alpha}_{\pm}\partial_x \tilde{\alpha}_{\pm}~~.
\end{eqnarray}
According to the beautiful explanation given in \cite{P}
the general solution for this equation of motion is described by
the deviation of the Fermi surface from  static ground-states.
The small deviation from the $\mu=0$ static ground-state
leads  functional scattering equations (\ref{71}).

      Nextly we shall consider   another classical equations (\ref{59})
which are obtained from  (\ref{56}) and ({\ref{75}})
under
$\hbar = -1/(i \mu)  \rightarrow 0$ limit.
We can follow the same steps as for the first case.
Equations (\ref{59}) get the following bosonized forms
\begin{eqnarray}
\alpha_{-2n}^{(+)} &=&
\sum_{r \geq 1}\frac{\Gamma(1+2n)}{\Gamma(2+2n-r) r!}
\sum_{m_1,\cdots,m_{r}}
\delta_{2n+\sum_{j=1}^{r}m_j, 0}
\alpha_{m_1}^{(-)}\cdots \alpha_{m_{r}}^{(-)}~~, \nonumber \\
\alpha_{2n}^{(-)} &=&
\sum_{r \geq 1}\frac{\Gamma(1+2n)}{\Gamma(2+2n-r) r!}
\sum_{m_1,\cdots,m_{r}}
\delta_{-2n+\sum_{j=1}^{r}m_j, 0}
\alpha_{m_1}^{(+)}\cdots \alpha_{m_{r}}^{(+)}~~.
\label{68}
\end{eqnarray}
These equations are also expected to hold for the noncompact case
with the similar modification as in (\ref{691})
\begin{eqnarray}
\alpha_{q}^{(\pm)} &=&
\sum_{r \geq 1}
\frac{\Gamma(1\mp q)}{\Gamma(2 \mp q-r)r!}
\int \prod_{j=1}^{r}dp_j
\delta(-q+\sum_{j=1}^{r}p_j)
\alpha_{p_1}^{(\mp)}\cdots \alpha_{p_{r}}^{(\mp)} ~~.
\label{681}
\end{eqnarray}
These are   formal solutions for
another variation of  Polchinski's
scattering equations
\begin{eqnarray}
\alpha_{\pm}(y)~=~\alpha_{\mp}(y \mp \ln (1+ \alpha_{\pm}(y))) ~~,
\label{70}
\end{eqnarray}
which were derived in \cite{P,MP} from the collective field study of
the classical
scattering in the undeformed $c=1$ string model.

          After the quantization of these (collective) boson fields
$\alpha_{\pm}(y)$,
their modes  $\alpha^{(+)}_n$ and $\alpha^{(-)}_{-n}$
($n>0$) ,
multiplied by appropriate leg factors, can be considered
as the creation operators of the massless tachyons with momentum
$n$ and $-n$ respectively \cite{JY,MP}.
Since  classically $\alpha^{(+)}_{n} = P_{+}^{~n}$
and $\alpha^{(-)}_{-n}=P_-^{~-n}$
in our context,
the quantum analogue of this identification is
$\alpha^{(+)}_n=W_n^{(0)}$ and $\alpha^{(-)}_{-n}=
g^{-1}W^{(0)}_{-n}g$, that is,
$\alpha^{(\pm)}_n=W_n^{(0) (\pm)}$.
Hence the free boson fields $\partial \phi_{\pm}$
which are introduced for  realizations (\ref{bosonrep})
of $W_{1+\infty}^{(\pm)}$ are identified with these
quantized collective fields  $\alpha_{\pm}$.
With this identification it follows that
the nonperturbative analogues of  formal solutions
(\ref{691}) and (\ref{681})
of  Polchinski's classical scattering equations  are
equations (\ref{742}) realized by $\partial \phi_{\pm}$.
Equivalently,   intertwining relation (\ref{quantumscattering})
of $W_{1+\infty}^{(\pm)}$  can be considered as
the nonperturbative form of
Polchinski's classical scattering equations.
We should also notice that, under  correspondence
(\ref{48}),
these modes of the $U(1)$ current,
$W^{(0) (\pm)}(z)$, give the asymptotic operators
$B_{\frac{n}{2}}^{in}$ and   $g^{-1}\bar{B}_{\frac{n}{2}}^{out}g$.
Therefore, taking it in reverse order,
one can say that
$B_n^{in}$ and $\bar{B}_n^{out}$, the asymptotic forms of
$B_n$ and $\bar{B}_n$, act as the creation operators
of the massless tachyons
with momentum $2n$ and $-2n$ respectively.
In paticular, the nonperturbative scatterings of the
(rescaled) massless
tachyons in this compactified string
will be  described by the  matrix elements
\begin{eqnarray}
<in|\prod_{i}B_{n_i}^{in}\prod_{j}\bar{B}^{out}_{m_j}|out>~~~,
\end{eqnarray}
which become equivalent to
\begin{eqnarray}
<0|\prod_{i}W_{2n_i}^{(0)}g \prod_j W_{-2m_j}^{(0)}|0>~~~.
\end{eqnarray}
So the generating function of these nonperturbative tachyon
scattering amplitudes, $\cal{F}$, are given by
\begin{eqnarray}
e^{\mbox{$\cal{F}$}(t,\bar{t})}=
 <0|e^{\sum_{k \geq 1}t_kW_k^{(0)}}
ge^{-\sum_{k \geq 1}\bar{t}_kW_{-k}^{(0)}}|0>.
\label{partition}
\end{eqnarray}
This partition function can be also considered as a $\tau$ function of
the Toda lattice hierarchy, which makes it possible to investigate
the tachyon scattering amplitudes from the view of integrable system.
It is a  main topic in the next section.

                The classical scattering amplitudes of the tachyons
of the  $c=1$  string in the black-hole background
were described in \cite{JY,DR,DK} as
the tree-level scattering amplitudes among these
collective modes $\alpha_n^{(\pm)}$,
which  are obtained  by utilizing
the quantum analogue of  relations
(\ref{691}).
Relations (\ref{691}) themself imply
the data of the black-hole background.
In these calculations, the Fermi energy $\mu$ was put zero from
the beginning.
But  classical relations
(\ref{691}) are
applicable even for the case of $\mu \neq  0$ due to the fact that
the contribution of $\mu$ vanishes, by the dimensional reason,
under the classical limit $\hbar =-1/i \sqrt{m}  \rightarrow 0$.
Therefore, at least,
these classical scattering amplitudes do not alter
under the shift of $\mu$.
The effect of $\mu$ will appear at the string multi-loops, that is,
the higher order contribution of the asymptotic expansion by
$\hbar=-1/(i \sqrt{m})$.
One may observe it
from  (\ref{88}),
in which $\mu$ never arise alone, but only in such a form as
$\mu \hbar$.
This explains why $\mu$ appears only in the multi-loop amplitudes.
The same reasoning  does hold for the case
of the classical scattering amplitudes
of the tachyons in the flat background.
These amplitudes was given \cite{MP} \cite{DJ,DJR} by
the tree-level amplitudes
among the collective modes
which are obtainable making use of the
quantum analogue of  (\ref{681}).
These equations (\ref{681}) were derived  \cite{P,MP}
as the classical
result for the collective field  approach
with the inverse harmonic oscilator potential
($m=0$) but, as we have shown,
they  also  hold
for the case of $m \neq 0$.
Thus the classical scattering amplitudes
do not change under the shift of $m$.
The effect of $m$ will appear at the higher order contribution
of the asymptotic expansion by
$\hbar =-1/(i \mu)$. (see  equations (\ref{89}).)
These features will mean that, though we have two parameters
$\mu$ and $m$ in our nonrelativistic fermion system
(\ref{1}) which defines a critical theory
for nonperturbative $c=1$ strings , one of these parameters becomes
irrelevant at the classical string level
which is defined as the leading contribution of
the asymptotic expansion by the another
parameter.
In this sense,
though fermion system (\ref{1}) itself can be easily deformed
from the type I theory of $c=1$ string in the flat background
($m=0$) to that in the proposed black hole background
($\mu=0$) on the parameter space $(\mu,m)$,
this deformation  is very nonperturbative from the string theoretical
point of view.

%%%%%%%%%%%%%%%%%%%%%%%%%%%%%%%%%%%%%%%%%%%%%%%%%%%%%%%%%%%%%%%
%%%%%%%%%%%%%%%%%%%%%%%%%%%%%%%%%%%%%%%%%%%%%%%%%%%%%%%%%%%%%%%%

\section{Toda Lattice Hierarchy and Compactified $c = 1$ Strings}

               The goal of this section is
to demonstrate that the Toda lattice
hierarchy provides a clear interpretation of our observations on
compactified $c = 1$ strings.  We present below a rather detailed
overview on the theory of the Toda lattice hierarchy
\cite{bib:TT-review} in three different languages ---  difference
operators, infinite matrices and free fermions.  The $W_{1+\infty}$
algebra discussed in the last section will emerge in several
different forms in these three frameworks. We then return to
$c = 1$ strings, and attempt to reorganize the contents of the
last section in terms of the Toda lattice hierarchy.

Throughout this section, $t = (t_1,t_2,\ldots)$ and $\tbar =
(\tbar_1,\tbar_2,\ldots)$ denote the two sets of ``time''
variables in the Toda lattice hierarchy, and $q \in Z$ the
``lattice'' coordinate. Thus $q$ should not be confused with
the momentum index $q$ of fermion modes in the preceding
sections. One will however notice that they actually play
the same role.

%%%%%%%%%%%%%%%%%%%%%%%%%%%%%%%%%%%%%%%%%%%%%%%%%%%%%%%%%%%%%%%%
\subsection{Lax formalism, dressing operators and $\tau$ function}

The Lax formalism of the Toda lattice hierarchy is based on the
following two Lax operators $L$ and $\Lbar$ and the Orlov-Shulman
operators $M$ and $\Mbar$.
\beqnarray
    L &=& e^{\rd_q} + \sum_{n=0}^\infty u_{n+1} e^{-n\rd_q},
                                                    \nonumber \\
    M &=& \sum_{n=1}^\infty n t_n L^n + q
        + \sum_{n=1}^\infty v_n L^{-n},
                                                    \nonumber \\
    \Lbar &=& \utilde_0 e^{\rd_q}
            + \sum_{n=0}^\infty \utilde_{n+1} e^{(n+2)\rd_q},
                                                    \nonumber \\
    \Mbar &=& - \sum_{n=1}^\infty n \tbar_n \Lbar^{-n} + q
            + \sum_{n=1}^\infty \vbar_n \Lbar^n,
\eeqnarray
where $e^{n\rd_q}$ denotes the shift operator that acts on a
function of $q$ as $ e^{n\rd_q} f(q) = f(q + n)$.  The above
operators are formal linear combinations of these shift operators,
and like pseudo-differential operators in the Lax formalism of
the KP hierarchy, these ``pseudo-difference'' operators form
a well defined non-commutative algebra. The coefficients $u_n$,
$v_n$, $\utilde_n$ and $\vbar_n$ are functions of $(t,\tbar,q)$,
$u_n = u_n(t,\tbar,q)$, etc.  We shall frequently write $u_n = u_n(q)$
etc., omitting $(t,\tbar)$, in order to stress the $q$ dependence.
The expansion of $L$ and $\Lbar$ appears somewhat asymmetric; it is
rather $\Lbar^{-1}$ that should be considered a counterpart of $L$.
Let $\ubar_n$ denote coefficients of the expansion of $\Lbar$:
\beqn
    \Lbar^{-1}~ =~ \ubar_0 e^{-\rd_q}
                + \sum_{n=0}^\infty \ubar_{n+1} e^{n\rd_q}~~.
\eeqn

Those Lax-Orlov-Shulman operators are required to satisfy the
twisted canonical commutation relations
\beqn
    [~L~,~ M~]= L~,~ \quad [~\Lbar~,~ \Mbar~]= \Lbar
                                        \label{eq:Toda-twistedCCR}
\eeqn
and the Lax equations
\beqnarray
    \frac{\rd L}{\rd t_n} = [A_n, L], &\quad&
    \frac{\rd L}{\rd \tbar_n} = [\Abar_n, L],
                                              \nonumber \\
    \frac{\rd M}{\rd t_n} = [A_n, M], &\quad&
    \frac{\rd M}{\rd \tbar_n} = [\Abar_n, M],
                                              \nonumber \\
    \frac{\rd \Lbar}{\rd t_n} = [A_n, \Lbar], &\quad&
    \frac{\rd \Lbar}{\rd \tbar_n} = [\Abar_n, \Lbar],
                                              \nonumber \\
    \frac{\rd \Mbar}{\rd t_n} = [A_n, \Mbar], &\quad&
    \frac{\rd \Mbar}{\rd \tbar_n} = [\Abar_n, \Mbar],
\eeqnarray
where $A_n$ and $\Abar_n$ are given by
\beqn
    A_n = ( L^n )_{\ge 0}, \quad
    \Abar_n = ( \Lbar^{-n} )_{< 0},
\eeqn
and $(\quad)_{\ge 0, <0}$ denotes the projection
\beqn
    ( \sum_n a_n e^{n \rd_q} )_{\ge 0} = \sum_{n \ge 0} a_n e^{n\rd_q},
    \quad
    ( \sum_n a_n e^{n \rd_q} )_{< 0} = \sum_{n < 0} a_n e^{n\rd_q}.
\eeqn
We call (\ref{eq:Toda-twistedCCR}) ``twisted'' because it is
rather the twisted operators $ML^{-1}$ and $\Mbar\Lbar^{-1}$ that
give canonical conjugate variable of $L$ and $\Lbar$:
\beqn
    [~L~,~ ML^{-1}~] = 1~,~ \quad [~\Lbar~,~ \Mbar\Lbar^{-1}~] = 1~~~.
\eeqn
The same kind of twisting will be used in the derivation of
string equations.

The above equations for the Lax-Orlov-Shulman operators,
as in the case of the KP hierarchy, can be converted into
equations for dressing operators. The Toda lattice hierarchy
needs two dressing operators
\beqnarray
    W = 1 + \sum_{n=1}^\infty w_n e^{-n\rd_q},
    &\quad&
    w_n = w_n(t,\tbar,q),
                                              \nonumber \\
    \Wbar = \wbar_0 + \sum_{n=1}^\infty \wbar_n e^{n\rd_q},
    &\quad&
    \wbar_n = \wbar_n(t,\tbar,q)
\eeqnarray
because of the presence of two different types of Lax-Orlov-Shulman
operators.  Twisted canonical commutation relations
(\ref{eq:Toda-twistedCCR}) are automatically satisfied if $L$, $M$,
$\Lbar$ and $\Mbar$ are written
\beqnarray
   L = W e^{\rd_q} W^{-1}, &\quad&
   M = W ( q + \sum_{n=1}^\infty n t_n e^{n\rd_q} ) W^{-1},
                                              \nonumber \\
   \Lbar = \Wbar e^{\rd_q} \Wbar^{-1}, &\quad&
   \Mbar = \Wbar ( q - \sum_{n=1}^\infty n \tbar_n e^{-n\rd_q} )
           \Wbar^{-1}.
                                      \label{eq:Toda-dressingRel}
\eeqnarray
This does not determine $W$ and $\Wbar$ uniquely.  By virtue of
this arbitrariness, one can single out a suitable pair of
$W$ and $\Wbar$ in such a way that they satisfy the equations
\beqnarray
    \dfrac{\rd W}{\rd t_n} = A_n W - W e^{n \rd_q}, &\quad&
    \dfrac{\rd W}{\rd \tbar_n} = \Abar_n W,
                                                    \nonumber \\
    \dfrac{\rd \Wbar}{\rd t_n} = A_n \Wbar,         &\quad&
    \dfrac{\rd \Wbar}{\rd \tbar_n}
                      = \Abar_n \Wbar - \Wbar e^{-n \rd_q}.
\eeqnarray
In fact, one can rewrite $A_n$ and $\Abar_n$ in terms of $W$
and $\Wbar$ by inserting the above expression of $L$ and $\Lbar$
into the definition of $A_n$ and $\Abar_n$.  Thus the above
equations can also be written
\beqnarray
    \dfrac{\rd W}{\rd t_n}
      = - (W e^{n\rd_q} W^{-1})_{<0} W,        &&
    \dfrac{\rd W}{\rd \tbar_n}
      = (\Wbar e^{-n \rd_q} \Wbar^{-1})_{<0} W,
                                               \nonumber \\
    \dfrac{\rd \Wbar}{\rd t_n}
      = (W e^{n \rd_q} W^{-1})_{\ge 0} \Wbar,  &&
    \dfrac{\rd \Wbar}{\rd \tbar_n}
      = - (\Wbar e^{-n \rd_q} \Wbar^{-1})_{\ge 0} \Wbar.
                                       \label{eq:Toda-WWbarFlowEq}
\eeqnarray
Note that this gives a closed system of equations defining commuting
flows in the space of dressing operators.

We can now introduce the notion of Baker-Akhiezer functions and
$\tau$ function of the Toda lattice hierarchy. The Baker-Akhiezer
functions are functions of $(t,\tbar,q)$ and $z$
(``spectral parameter'') of the form
\beqnarray
   \Psi &=& ( 1 + \sum_{n=1}^\infty w_n z^{-n} )
          z^q \exp( \sum_{n=1}^\infty t_n z^n ),
                                         \nonumber \\
   \Psibar &=& ( \wbar_0 + \sum_{n=1}^\infty \wbar_n z^n )
             z^q \exp( \sum_{n=1}^\infty \tbar_n z^{-n} ),
\eeqnarray
and satisfy a system of linear equations, and its integrability
condition is exactly the Lax equations and the twisted canonical
commutation relations above.  The $\tau$ function
$\tau = \tau(t,\tbar,q)$, by definition, is a function that is
connected with the Baker-Akhiezer functions as:
\beqnarray
    \Psi &=&
      \dfrac{ \exp \left( -\sum_{n=1}^\infty \frac{z^{-n}}{n}
              \frac{\rd}{\rd t_n} \right) \tau(t,\tbar,q) }
            { \tau(t,\tbar,q) }
      z^q \exp( \sum_{n=1}^\infty t_n z^n ),
                                             \nonumber \\
    \Psibar &=&
      \dfrac{ \exp \left( -\sum_{n=1}^\infty \frac{z^n}{n}
              \frac{\rd}{\rd \tbar_n} \right) \tau(t,\tbar,q+1) }
            { \tau(t,\tbar,q) }
      z^q \exp( \sum_{n=1}^\infty \tbar_n z^{-n} ).
\eeqnarray
If expanded into Laurent series of $z$, these relations give
a system of linear differential equations for $\tau$, whose
integrability condition is now ensured by the linear equations
of the Baker-Akhiezer functions or, equivalently, by equations
(\ref{eq:Toda-WWbarFlowEq}) of commuting flows of the dressing
operators.

       A few historical remarks will be now in order. The Toda lattice
hierarchy was first developed as a discrete (or difference)
version of the KP hierarchy \cite{bib:UT-Toda}. Fundamental
tools such as the dressing operators, the Baker-Akhiezer
functions, the tau function, etc.,  were simultaneously
imported from the theory of the KP hierarchy at that stage
\cite{bib:Sato-Sato,bib:DJKM,bib:Segal-Wilson}. (Our present
notations, however, are considerably different from notations
in these earlier works.) Meanwhile, the notion of the Orlov-Shulman
operators appeared several years later \cite{bib:Orlov-etal}, and
extended to the Toda lattice hierarchy rather recently
\cite{bib:TT-Toda,bib:ASvM}. Orlov and his collaborators used
such an operator to describe the so called ``additional symmetries''
(such as Virasoro symmetries) of integrable hierarchies within the
Lax formalism.

%%%%%%%%%%%%%%%%%%%%%%%%%%%%%%%%%%%%%%%%%%%%%%%%%%%%%%%%%%%%%%%%%%
\subsection{Matrix representation and semi-infinite determinant}

The Toda lattice hierarchy can also be formulated in the language
of infinite matrices. This is due to the following correspondence
between difference operators and infinite ($Z \times Z$) matrices:
\beqn
    \sum_n a_n(q) e^{n\rd_q}
    \longleftrightarrow
    \sum_n \diag[a_n(i)] \Lambda^n
\eeqn
where
\beqn
    \diag[a_n(i)] = \Bigl( a_n(i) \delta_{ij} \Bigr)~,~
    \quad
    \Lambda^n = \Bigl( \delta_{i,j-n} \Bigr)~~,
\eeqn
and the indices $i$ (row) and $j$ (column) run over $Z$.
With this correspondence, algebraic operations for difference
operators are mapped to the corresponding operations for infinite
matrices. Similarly, the projection $(\quad)_{\ge 0,<0}$ becomes
the projection onto upper triangular part (including the diagonal)
and lower triangular part (excluding the diagonal) of infinite
matrices:
\beqnarray
    ( A )_{\ge 0} &=& \Bigl( \theta(j-i) a_{ij} \Bigr),
                                             \nonumber \\
    ( A )_{< 0} &=& \Bigl( (1 - \theta(j-i)) a_{ij} \Bigr).
\eeqnarray
Accordingly, the Lax, Orlov-Shulman and dressing operators
have the corresponding infinite matrices
\beqn
    L     \leftrightarrow \bL,    \quad
    M     \leftrightarrow \bM,    \quad
    \Lbar \leftrightarrow \bLbar, \quad
    \Mbar \leftrightarrow \bMbar, \quad
    W     \leftrightarrow  \bW,    \quad
    \Wbar \leftrightarrow \bWbar,
\eeqn
and those matrices obey the same form of equations as we have
presented in the previous subsection.  Note, in particular, that
dressing representation (\ref{eq:Toda-dressingRel}) of the Lax
and Orlov-Shulman operators turns into matrix relations
of the form
\beqnarray
   \bL = \bW \Lambda \bW^{-1}, &\quad&
   \bM = \bW ( \Delta + \sum_{n=1}^\infty n t_n \Lambda^n )
         \bW^{-1},
                                              \nonumber \\
   \bLbar = \bWbar \Lambda \bWbar^{-1}, &\quad&
   \bMbar = \bWbar ( \Delta - \sum_{n=1}^\infty n \tbar_n \Lambda^{-n} )
            \bWbar^{-1}.
\eeqnarray
Here $\Delta$ is the infinite matrix
\beqn
    \Delta = \Bigl( i \delta_{ij} \Bigr)
\eeqn
that represents the multiplication operator $q$.
The infinite matrices $\bW$ and $\bWbar$ are
lower and upper triangular matrices of the form
\beqnarray
    \bW &=&
      \left( \begin{array}{ccccc}
        \ddots &         &        &        &         \\
        \ddots & 1       &        & 0      &         \\
        \ddots & w_1(-1) & 1      &        &         \\
        \ddots & \ddots  & w_1(0) & 1      &         \\
        \ddots & \ddots  & \ddots & \ddots & \ddots
      \end{array} \right),
                                             \nonumber \\
    \bWbar &=&
      \left( \begin{array}{ccccc}
        \ddots & \ddots      & \ddots      & \ddots     & \ddots \\
               & \wbar_0(-1) & \wbar_1(-1) & \ddots     & \ddots \\
               &             & \wbar_0(0)  & \wbar_1(0) & \ddots \\
               & 0          &             & \ddots     & \ddots \\
               &             &             &            & \ddots
      \end{array} \right).
                                     \label{eq:Toda-MatrixWWbar}
\eeqnarray

Having this reformulation of the Toda lattice hierarchy, we now show
that solving the Toda lattice hierarchy can be reduced to a problem
of linear algebra (of $Z \times Z$ matrices). A key role is played
by the ``matrix ratio'' $\bU(t,\tbar)$ of
$\bW = \bW(t,\tbar)$ and $\bWbar = \bWbar(t,\tbar)$:
\beqn
    \bU(t,\tbar) = \bW(t,\tbar)^{-1} \bWbar(t,\tbar).
                                       \label{eq:Toda-GaussDecom}
\eeqn
It is not hard to show from the matrix counterpart of
(\ref{eq:Toda-WWbarFlowEq}) that $\bU(t,\tbar)$ satisfies
the simple linear equations
\beqn
    \frac{\rd \bU(t,\tbar)}{\rd t_n} = \Lambda^n \bU(t,\tbar),
    \quad
    \frac{\rd \bU(t,\tbar)}{\rd \tbar_n} = - \bU(t,\tbar) \Lambda^{-n}.
                                      \label{eq:Toda-UFlowEq}
\eeqn
In other words, the complicated nonlinearity of the Toda lattice
hierarchy is ``linearized'' in $\bU(t,\tbar)$:
\beqn
    \bU(t,\tbar) = \exp( \sum_{n=1}^\infty t_n \Lambda^n ) \bU(0,0)
                   \exp( -\sum_{n=1}^\infty \tbar_n \Lambda^{-n} ).
\eeqn

What is more important is that this process is reversible.
Namely, if an infinite matrix $\bU(t,\tbar)$ of this form is
given, and if one can find such a pair of triangular infinite
matrices $\bW(t,\tbar)$ and $\bWbar(t,\tbar)$ of the form
(\ref{eq:Toda-MatrixWWbar}) that satisfy
(\ref{eq:Toda-GaussDecom}), then $\bW(t,\tbar)$ and
$\bWbar(t,\tbar)$ turn out to obey (\ref{eq:Toda-WWbarFlowEq}),
i.e., give a solution of the Toda lattice hierarchy.  To see this,
differentiate the both hand sides of (\ref{eq:Toda-GaussDecom})
with respect to $t$ and $\tbar$, rewrite the derivatives of
$\bU(t,\tbar)$ by (\ref{eq:Toda-UFlowEq}), and finally eliminate
$\bU(t,\tbar)$ using (\ref{eq:Toda-GaussDecom}). One will then
obtain
\beqnarray
       \frac{\rd \bW}{\rd t_n} \bW^{-1}
       + \bW \Lambda^n \bW^{-1}
   &=& \frac{\rd \bWbar}{\rd t_n} \bWbar^{-1},
                                                 \nonumber \\
       \frac{\rd \bW}{\rd \tbar_n} \bW^{-1}
       - \bWbar \Lambda^{-n} \bWbar^{-1}
   &=& \frac{\rd \bWbar}{\rd \tbar_n} \bWbar^{-1}.
\eeqnarray
The $(\quad)_{<0}$ part of these equations gives
\beqn
    \frac{\rd \bW}{\rd t_n} \bW^{-1}
      = - ( \bW \Lambda^n \bW^{-1} )_{<0},
                                           \quad
    \frac{\rd \bW}{\rd \tbar_n} \bW^{-1}
      =   ( \bWbar \Lambda^{-n} \bWbar^{-1} )_{<0},
\eeqn
which is the first two equations of (\ref{eq:Toda-WWbarFlowEq}).
The $(\quad)_{\ge 0}$ part, similarly, becomes
\beqn
    \frac{\rd \bWbar}{\rd t_n} \bWbar^{-1}
      =   ( \bW \Lambda^n \bW^{-1} )_{\ge 0},
                                           \quad
    \frac{\rd \bWbar}{\rd \tbar_n} \bWbar^{-1}
      = - ( \bWbar \Lambda^{-n} \bWbar^{-1} )_{\ge 0},
\eeqn
and gives the other two of (\ref{eq:Toda-WWbarFlowEq}).

This kind of solution technique is called a ``factorization method''
or a ``Riemann-Hilbert problem'' in the terminology of nonlinear
integrable systems \cite{bib:UT-Toda}. (The KP hierarchy, too,
can be treated by a similar factorization method \cite{bib:Mulase}.)
In the present case, factorization problem (\ref{eq:Toda-GaussDecom})
is a $GL(\infty)$ version of the Gauss decomposition in ordinary
finite dimensional linear algebra. At least formally, therefore,
one will be able to apply a standard method using the Cramer formula
to obtain the two factors $\bW$ and $\bWbar$ explicitly. Their
matrix elements $w_n(q)$ and $\wbar_n(q)$ can be written as a
quotient of two determinants. This calculation eventually leads
to the following formula for the $\tau$ function:
\beqn
    \tau(t,\tbar,q) = \det\Bigl( u_{ij}(t,\tbar)(i,j<q) \Bigr),
\eeqn
where $u_{ij}(t,\tbar)$ are matrix elements of $\bU(t,\tbar)$.
This determinant, as well as those emerging in the determinant
formulas of $w_n(q)$ and $\wbar_n(q)$, is a semi-infinite
determinant, and needs some justification. Fortunately, these
semi-infinite determinants are known to be well defined under
a suitable interpretation \cite{bib:T-TodaIni}.

%%%%%%%%%%%%%%%%%%%%%%%%%%%%%%%%%%%%%%%%%%%%%%%%%%%%%%%%%%%%%%%%%%%%%
\subsection{Symmetries and constraints}

Another implication of (\ref{eq:Toda-GaussDecom}) is the existence
of underlying $GL(\infty)$ and $W_{1+\infty}$ symmetries.  One can
use the language of $W_{1+\infty}$ symmetries to formulate string
equations in a general form.

$GL(\infty)$ symmetries are symmetries on the space of solutions of
the Toda lattice hierarchy, and induced by the left and right action
of $GL(\infty)$ on $\bU = \bU(0,0)$.  Thus, more precisely, they
are $GL(\infty) \times GL(\infty)$ symmetries. Since we are mostly
interested in their infinitesimal form, let us consider a
one-parameter family of such deformations,
\beqn
   \bU \to \bU(\epsilon)
     = e^{-\epsilon \bA} \bU e^{\epsilon \bAbar},
\eeqn
where $\bA = (a_{ij})$ and $\bAbar = (\abar_{ij})$ are arbitrary
$Z \times Z$ matrices (i.e., elements of $gl(\infty)$).
By Gauss decomposition (\ref{eq:Toda-GaussDecom}), this should
induce a one-parameter family of deformations of the dressing
operators $\bW$ and $\bWbar$. As we shall show below, one can
easily derive differential equations satisfied by $\bW$ and
$\bWbar$ with respect to the deformation parameter.  These
differential equations give an infinitesimal expression of the
$GL(\infty)$ symmetries. (This kind of methods for constructing
finite and infinitesimal symmetries of nonlinear integrable
systems have been used for years, and called ``Riemann-Hilbert
transformations'', ``dressing transformations'', etc.
The formulation presented here is borrowed from a similar
formulation in the self-dual Yang-Mills equations and the
self-dual Einstein equation \cite{bib:T-SDEq}.)

Let $\bW(\epsilon) = \bW(\epsilon,t,\tbar)$ and
$\bWbar(\epsilon) = \bWbar(\epsilon,t,\tbar)$ denote
the corresponding deformations of $\bW$ and $\bWbar$.
They are characterized by the Gauss decomposition
\beqn
    \bU(\epsilon,t,\tbar)
    = \bW(\epsilon,t,\tbar)^{-1} \bWbar(\epsilon,t,\tbar),
\eeqn
where
\beqn
    \bU(\epsilon,t,\tbar)
    = \exp( \sum_{n=1}^\infty t_n \Lambda^n ) \bU(\epsilon)
      \exp( - \sum_{n=1}^\infty \tbar_n \Lambda^{-n} ).
\eeqn
One can now repeat almost the same calculations as we derived
(\ref{eq:Toda-WWbarFlowEq}) from (\ref{eq:Toda-GaussDecom})
to obtain the following equations with respect to the
deformation parameter $\epsilon$:
\beqnarray
    \frac{\rd \bW(\epsilon)}{\rd \epsilon}
    &=& \Bigl(
            \bW(\epsilon) \bA(t) \bW(\epsilon)^{-1}
          - \bWbar(\epsilon) \bAbar(\tbar) \bWbar(\epsilon)
        \Bigr)_{<0} \bW(\epsilon),
                                                   \nonumber \\
    \frac{\rd \bWbar(\epsilon)}{\rd \epsilon}
    &=& \Bigl(
            \bWbar(\epsilon) \bAbar(\tbar) \bWbar(\epsilon)
          - \bW(\epsilon) \bA(t) \bW(\epsilon)^{-1}
        \Bigr)_{\ge 0} \bWbar(\epsilon).
\eeqnarray
Here $\bA(t)$ and $\bAbar(\tbar)$ are given by
\beqnarray
    \bA(t)
      &=& \exp( \sum_{n=1}^\infty t_n \Lambda^n ) \bA
          \exp( - \sum_{n=1}^\infty t_n \Lambda^n),
                                                   \nonumber \\
    \bAbar(\tbar)
      &=& \exp( \sum_{n=1}^\infty \tbar_n \Lambda^{-n} ) \bAbar
          \exp( - \sum_{n=1}^\infty \tbar_n \Lambda^{-n} ).
\eeqnarray

We can thus obtain the following expression of an infnitesimal
$GL(\infty) \times GL(\infty)$ symmetry:
\beqnarray
    \delta_{\bA,\bAbar} \bW
    &=& \Bigl( \bW \bA(t) \bW^{-1} - \bWbar \bAbar(\tbar) \bWbar^{-1}
        \Bigr)_{<0} \bW,
                                                   \nonumber \\
    \delta_{\bA,\bAbar} \bWbar
    &=& \Bigl( \bWbar \bAbar(\tbar) \bWbar^{-1} - \bW \bA(t) \bW^{-1}
        \Bigr)_{\ge 0} \bWbar.
                                    \label{eq:Toda-deltaWWbar}
\eeqnarray
Viewing $\delta_{\bA,\bAbar}$ as an abstract variational operator
acting on functionals of dressing operators, one can derive the
commutation relations
\beqn
    [ \delta_{\bA_1,\bAbar_1}, \delta_{\bA_2,\bAbar_2} ]
    = \delta_{[\bA_1,\bA_2],[\bA_2,\bAbar_2]}.
\eeqn
In particular, $\delta_{\bA,0}$ and $\delta_{0,\bAbar}$ give
two independent $GL(\infty)$ symmetries that commute with
each other.  Of course this is to be expected from their
origin as left and right actions of $GL(\infty)$ on
the matrix $\bU$.

$W_{1+\infty}$ symmetries (more precisely, $W_{1+\infty} \oplus
W_{1+\infty}$ symmetries) are a subset of these $GL(\infty)$
symmetries with generators of the special form
\beqnarray
    \bA ~=~ A(\Delta,\Lambda)
    &=& \sum_{m,\ell} a_{m\ell} \Delta^m \Lambda^\ell,
                                                   \nonumber \\
    \bAbar ~=~ \Abar(\Delta,\Lambda)
    &=& \sum_{m,\ell} \abar_{m\ell} \Delta^m \Lambda^\ell,
\eeqnarray
where $m$ runs over nonnegative integers and $\ell$ over all
integers.  Note that these matrices are in one-to-one correspondence
with difference operators,
\beqn
    \bA(\Delta,\Lambda) \leftrightarrow \bA(q,e^{\rd_q}), \quad
    \bAbar(\Delta,\Lambda) \leftrightarrow \bAbar(q,e^{\rd_q}),
\eeqn
and altogether form a closed Lie algebra with the fundamental
commutation relation
\beqn
    [~\Delta~,~ \Lambda~] = \Delta \ \leftrightarrow \
    [~e^{\rd_q}~,~ q~] = e^{\rd_q}.
\eeqn
This is essentially a (centerless) $W_{1+\infty}$ algebra,
the generators $W^{(k)}_n$ being given by
\beqn
    W^{(k)}_n = (\Delta\Lambda^{-1})^k \Lambda^{n+k}
              ~\sim~ (q e^{-\rd_q})^k e^{(n+k)\rd_q}.
\eeqn
One can easily see that for such $\bA$ and $\bAbar$,
previous formula (\ref{eq:Toda-deltaWWbar}) becomes
\beqnarray
   \delta_{\bA,\bAbar} \bW
   &=& \Bigl( A(\bM,\bL) - \Abar(\bMbar,\bLbar)
       \Bigr)_{<0} \bW,
                                                \nonumber \\
   \delta_{\bA,\bAbar} \bWbar
   &=& \Bigl( \Abar(\bMbar,\bLbar) - A(\bM,\bL)
       \Bigr)_{\ge 0} \bWbar,
\eeqnarray
where
\beqn
    A(\bM,\bL)
      = \sum_{m,\ell} a_{m\ell} \bM^m \bL^\ell,  \quad
    \Abar(\bMbar,\bLbar)
      = \sum_{m,\ell} \abar_{m\ell} \bMbar^m \bLbar^\ell.
\eeqn
The left piece $\delta_{\cA,0}$ and the right piece
$\delta_{0,\cAbar}$, as in the case of $GL(\infty)$ above,
give two independent sets of $W_{1+\infty}$ symmetries
commuting with each other, i.e., a realization of
$W_{1+\infty} \oplus W_{1+\infty}$.

Furthermore, these symmetries turn out to induce the following
symmetries on the Lax-Orlov-Shulman operators:
\beqnarray
    \delta_{\bA,\bAbar} \bL
    &=& \Bigl[ \Bigl( A(\bM,\bL) - \Abar(\bMbar,\bLbar)
               \Bigr)_{<0}, \bL \Bigr],
                                                  \nonumber \\
    \delta_{\bA,\bAbar} \bM
    &=& \Bigl[ \Bigl( A(\bM,\bL) - \Abar(\bMbar,\bLbar)
               \Bigr)_{<0}, \bM \Bigr],
                                                  \nonumber \\
    \delta_{\bA,\bAbar} \bLbar
    &=& \Bigl[ \Bigl( \Abar(\bMbar,\bLbar) - A(\bM,\bL)
               \Bigr)_{\ge 0}, \bLbar \Bigr],
                                                  \nonumber \\
    \delta_{\bA,\bAbar} \bMbar
    &=& \Bigl[ \Bigl( \Abar(\bMbar,\bLbar) - A(\bM,\bL)
               \Bigr)_{\ge 0}, \bMbar \Bigr].
\eeqnarray
Note that these relations are closed within the Lax-Orlov-Shulman
operators --- the dressing operators have disappeared.  This fact
lies in the heart of our understanding of string equations.

Because of this, we can formulate a $W_{1+\infty}$ constraint
as a constraint on the Lax-Orlov-Shulman operators. In general,
such a $W_{1+\infty}$ constraint can be formulated as the fixed
point condition $\bU(\epsilon) = \bU$, which is equivalent to
\beqn
    A(\Delta,\Lambda) \bU ~=~ \bU \Abar(\Delta,\Lambda)~~.
\eeqn
In terms of the Lax-Orlov-Shulman operators, this fixed point
condition can be restated as $\delta_{\bA,\bAbar}\bL = 0$, etc.,
or, equivalently, as
\beqn
    A(\bM,\bL) ~=~ \Abar(\bMbar,\bLbar)~~.
\eeqn
In actual applications, several constraints of this form
are considered simultaneously, frequently taking the form of
a ``$(P,Q)$ pair''. This type of equations are what we call
``string equations''.

One can further translate the above constraint into equations
of the form
\beqn
    X_\bA(t,\rd_t,q) \tau = \Xbar_\bAbar(\tbar,\rd_\tbar,q) \tau,
\eeqn
where $X_\bA(t,\rd_t,q)$ and $\Xbar_\bAbar(\tbar,\rd_\tbar,q)$
are linear differential operators in $t$ and $\tbar$ that
represent the action of $\delta_{\bA,0}$ and $\delta_{0,\bAbar}$
in terms of the $\tau$ function.  This is a general form of
the so called ``$W_{1+\infty}$-constraints'' in the literature
on $c = 1$ strings.  We shall specify the origin of these
differential operators $X_\bA(t,\rd_t,q)$ and
$\Xbar_\bAbar(\tbar,\rd_\tbar,q)$ in the next subsection.

These constraints may be thought of as imposing a relation
between the (otherwise independent) left and right components
of the full $W_{1+\infty} \oplus W_{1+\infty}$ structure.
We have indeed observed many aspects of such a relation in the
analysis of deformed $c = 1$ strings.  This issue, too, will
be discussed in detail below.

%%%%%%%%%%%%%%%%%%%%%%%%%%%%%%%%%%%%%%%%%%%%%%%%%%%%%%%%%%%%%%
\subsection{Free fermion formalism}

We are now in a position to reformulate the contents of the
preceding two subsections in terms of free fermions. Notations
concerning fermions, such as the fermion operators $\psi_n$,
$\psi^*_n$, etc., are the same as those used in the analysis
of compactified $c = 1$ strings.

To begin with, let us recall the following correspondence between
the $gl(\infty)$ algebra (more precisely,
its central extension $\hat{gl}(\infty)$)
and fermion bilinear forms:
\beqn
    \bA = (a_{ij}) ~\leftrightarrow~
    A(\psi,\psi^*) = \sum_{i,j} a_{ij} : \psi_{-i} \psi^*_j :~~.
\eeqn
Combining this correspondence with the aforementioned correspondence
between difference operators and infinite matrices, one can obtain a
free fermion realization of difference operators:
\beqn
    :q^m e^{\ell \rd_q}:
   ~ = ~\sum_{p=-\infty}^\infty p^m :\psi_{-p} \psi^*_{p+\ell}:
\eeqn
In particular, the $W_{1+\infty}$ generators $W^{(k)}_n$ are
now represented by
\beqn
    W^{(k)}_n ~=~ : (q e^{-\rd_q})^k e^{(n+k)\rd_q} :~~.
\eeqn

The next step is to lift up the above correspondence to the
group level.  A fermion bilinear form induces a linear map
on the vector space spanned by $\psi_n$ and $\psi^*_n$:
\beqn
    [~ A(\psi,\psi^*)~,~ \psi_{-j}~ ] = \sum_i a_{ij} \psi_{-i}~,~
    \quad
    [~ A(\psi,\psi^*)~,~ \psi^*_i ~ ] = - \sum_j a_{ij} \psi^*_j~~.
\eeqn
An operator of the form $g = \exp A(\psi,\psi^*)$, accordingly,
induces an invertible linear transformation in the same vector space:
\beqnarray
  &&  g \psi_{-j} g^{-1} = \sum_i u_{ij} \psi_{-i},
      \quad
      g \psi^*_i g^{-1}  =  \sum_j v_{ij} \psi^*_j,
                                                \nonumber \\
  &&  (v_{ij}) = (u_{ij})^{-1}.
                                      \label{eq:Toda-gLinTrans}
\eeqnarray
It is  this $Z \times Z$ matrix $\bU = (u_{ij})$ that we
now identify with the previous matrix $\bU = \bU(0,0)$ derived
in the matrix formalism of the Toda lattice hierarchy.

To this end, we first note that the special matrix element
$<q \mid g \mid q>$ of the operator $g$ can be rewritten as the
determinant of a semi-infinite submatrix of $\bU$.  Let us
write the $q$-th ground state as the fermi sea of a half
filled vacuum:
\beqn
    \mid q> ~=~ \psi_{-q+1} \psi_{-q+2} \cdots \mid -\infty>.
\eeqn
Repeated use of (\ref{eq:Toda-gLinTrans}) will then give
\beqnarray
    g \mid q> &=& \Bigl(\sum_i u_{i,q-1} \psi_{-i}\Bigr) g \mid q-1>
                                                   \nonumber \\
          &=& \Bigl(\sum_i u_{i,q-1} \psi_{-i}\Bigr)
              \Bigl(\sum_i u_{i,q-2} \psi_{-i}\Bigr) g \mid q-2>
                                                   \nonumber \\
          &=& \ldots
                                                   \nonumber \\
          &=& \Bigl(\sum_i u_{i,q-1} \psi_{-i}\Bigr)
              \Bigl(\sum_i u_{i,q-2} \psi_{-i}\Bigr)
              \cdots \mid -\infty>.
\eeqnarray
The last expression can be expanded into an infinite linear
combination of Fock states
\beqn
    \mid \{n_i(i<q)\}>
    ~=~ \psi_{-n_{q-1}} \psi_{-n_{q-2}} \cdots \mid -\infty>,
\eeqn
and the coefficients are given by semi-infinite determinants of
the form $\det\Bigl( u_{n_i,j} (i,j<q) \Bigr)$.  Pairing with
$<q \mid$ leaves only the term proportional to $\mid q>$
non-vanishing, thereby we obtain the formula
\beqn
    <q \mid g \mid q> ~=~ \det\Bigl( u_{ij} (i,j < q) \Bigr)~~.
\eeqn

The last formula gives the special value $\tau(0,0,q)$ of
the $\tau$ function. The $\tau$ function itself is given by the
same semi-infinite determinant with $\bU$ replaced by
$\bU(t,\tbar)$. In the free fermion formalism, this amounts
to replacing
\beqn
    g \to g(t,\tbar) ~= ~e^{H(t)} g e^{-\Hbar(\tbar)},
\eeqn
where
\beqnarray
    H(t) = \sum_{n=1}^\infty t_n J_n, \quad
      \Hbar(\tbar) = \sum_{n=1}^\infty \tbar_n J_{-n},
                                           \nonumber \\
   J_n = W^{(0)}_n
       = \sum_{p=-\infty}^\infty :\psi_{-p} \psi^*_{p+n}:.
\eeqnarray
The reason is that $J_n$ give a fermionic representation of
$e^{n\rd_q} \sim \Lambda^n$ (i.e.,  $J_n = :e^{n\rd_q}:$).
One can indeed easily check the following commutation
relations with $\psi_{-j}$ and $\psi^*_i$:
\beqn
    [~J_n~,~ \psi_{-j}~] = \psi_{-j+n},
    \quad
    [~J_n~, ~\psi^*_i~]  = \psi_{i+n}.
\eeqn
(Compare these relations with those of the previous
fermion bilinear operators $\cB_n$ and $\cBbar_n$.)
Accordingly $e^{H(t)}$ and $e^{\Hbar(\tbar)}$ give a
fermionic representation of $\exp(\sum t_n \Lambda^n)$
and $\exp(\sum \tbar_n \Lambda^{-n})$. We thus obtain the
following fermionic representation of the $\tau$ function
\cite{bib:JM-review,bib:Takebe-TodaTau}:
\beqn
    \tau(t,\tbar,q)~ =~ <q \mid e^{H(t)} g e^{-\Hbar(\tbar)} \mid q>.
\eeqn
This is actually a straightforward generalization of a similar
representation of the $\tau$ function of the KP hierarchy
\cite{bib:DJKM}.  In fact, one can readily see, from such a
comparison \cite{bib:UT-Toda}, that $\tau(t,\tbar,q)$ may be
viewed as a $\tau$ function of the KP hierarchy in two opposite
ways, namely
(i) $t$ as time variables and $(\tbar,q)$ as parameters, and
(ii) $\tbar$ as time variables and $(t,q)$ as parameters.

This fermionic representation of the $\tau$ function also shows
that the $\tau$ function is actually a generating function of
fermion scattering amplitudes discussed in the previous sections.
To demonstrate this, let us insert a complete set of Fock states,
$1 = \sum \mid\{n_i(i<q)\}><\{n_i(i<q)\}\mid$, between
$e^{H(t)}$ and $g$, and similarly, between $g$ and
$e^{-\Hbar(\tbar)}$.  This gives the expansion
\beqnarray
     \tau(t,\tbar,q)
     & = &\sum_{\{n_i\},\{\nbar_i\}}
        <q \mid e^{H(t)} \mid \{n_i(i<q)\}> \times
                                           \nonumber \\
     &&~~~\times
        <\{n_i(i<q)\} \mid g \mid \{\nbar_i(i<q)\}>
        <\{\nbar_i(i<q)\} \mid e^{-\Hbar(\tbar)} \mid q>~~.
\nonumber \\
{}~~~~~~
\eeqnarray
In fact, the matrix elements $<q \mid e^{H(t)} \mid \{n_i(i<q)\}>$
and $<\{\nbar_i(i<q)\}\mid e^{-\Hbar(\tbar)} \mid q>$ survive
only for such states as $n_i = i$ for $i << q$, and
coincide with the so called ``Shur functions''
\cite{bib:Sato-Sato,bib:DJKM,bib:Segal-Wilson}.  The matrix
elements of $g$ are compactified analogues of the scattering
amplitudes.

$W_{1+\infty}$ symmetries take a particularly simple form in
this fermionic representation of the $\tau$ function --- they
are given by inserting the corresponding fermion bilinear forms
to the left or right of $g$. Furthermore, as already mentioned
in the end of the last subsection, such a $W_{1+\infty}$ symmetry
can also be represented by a linear differential operator in $t$
or $\tbar$. These facts are summarized into the following generating
functional formulas \cite{bib:DJKM,bib:Takebe-TodaTau,bib:TT-Toda} :
\beqnarray
    <q \mid e^{H(t)} :\psi(z)\psi^*(w): g e^{-\Hbar(\tbar)} \mid q>
    &=& X(z,w) <q \mid e^{H(t)} g e^{-\Hbar(\tbar)} \mid q>,
                                               \nonumber \\
    <q\mid e^{H(t)} g :\psi(z)\psi^*(w): e^{-\Hbar(\tbar)} \mid q>
    &=& \Xbar(z,w) <q \mid e^{H(t)} g e^{-\Hbar(\tbar)} \mid q>,
                                               \nonumber \\
    &&
\eeqnarray
where $X(z,w)$ and $\Xbar(z,w)$ are given by
\beqnarray
    X(z,w) &=&
    \dfrac{ \exp\Bigl( \sum_{n=1}^\infty t_n (z^n - w^n) \Bigr)
            \Bigl(\frac{z}{w}\Bigr)^q
            \exp\Bigl( -\sum_{n=1}^\infty \frac{z^{-n} - w^{-n}}{n}
                                          \frac{\rd}{\rd t_n} \Bigr)
            - 1                                                    }
          { z - w },
                                               \nonumber \\
    \Xbar(z,w) &=&
    \dfrac{ \exp\Bigl( \sum_{n=1}^\infty \tbar_n (z^{-n} - w^{-n}) \Bigr)
            \Bigl(\frac{z}{w}\Bigr)^q
            \exp\Bigl( -\sum_{n=1}^\infty \frac{z^n - w^n}{n}
                                          \frac{\rd}{\rd \tbar_n} \Bigr)
            - 1                                                        }
          { z - w }.
                                                  \nonumber \\
    &&
\eeqnarray
An explicit form of the differential operators $X_\bA$ and
$\Xbar_\bAbar$ mentioned in the end of the last subsection
can be derived from these formulas. For instance, differential
operators representing the $W_{1+\infty}$ generators $W^{(k)}_n$
are given by
\beqnarray
    X^{(k)}_n &=&
      \oint \frac{dz}{2 \pi i} z^{n+k} \rd_z^k X(z,w)|_{w=z},
                                               \nonumber \\
    \Xbar^{(k)}_n &=&
      \oint \frac{dz}{2 \pi i} z^{n+k} \rd_z^k \Xbar(z,w)|_{w=z}.
\eeqnarray
These two sets of differential operators $X^{(k)}_n$ and
$\Xbar^{(k)}_n$ give generators of the two (left and right)
components in $W_{1+\infty} \oplus W_{1+\infty}$ symmetries
of the Toda lattice hierarchy.

The above correspondence between the fermion bilocal field and
a differential operator is just a disguised form of the well known
formula
\beqn
    :\psi(z) \psi^*(w):
    ~=~ \dfrac{ :e^{\phi(z)-\phi(w)}: - 1 }{z - w}
\eeqn
in the standard bosonization of relativistic free fermions,
$\psi(z) = :e^{\phi(z)}:$, $\psi^*(z) = :e^{-\phi(z)}:$.
This fact is particularly well understood in the case of the
KP hierarchy \cite{bib:DJKM}.
In that case, the boson (or $U(1)$ current) Fourier
modes $\alpha_n$ can be identified with differential operators
acting on the $\tau$ function as $\alpha_n \sim \rd / \rd t_n$
and $\alpha_{-n} \sim n t_n$. The right hand side of the above
bosonization formula of $:\psi(z)\psi^*(w):$ then reproduces the
operator $X(z,w)$.

A new feature in the Toda lattice hierarchy is that there are
two sets of time variables $t$ and $\tbar$.  This implies the
existence of two bosonizations with two boson fields, say
$\phi(z)$ and $\phibar(z)$. These boson fields correspond
to the two boson fields $\phi_\pm(z)$ of Section 3.  The
differential operators $X(z,w)$ and $\Xbar(z,w)$ may be
thought of as vertex operators in these two bosonization
schemes.  By Taylor expansion at $z = w$ and Fourier (Laurent)
expansion in $z$, one obtains two bosonic realizations
$X^{(k)}_n$ and $\Xbar^{(k)}_n$ of the $W_{1+\infty}$ algebra.

One can also see in the present framework that $g$ plays the role
of an intertwining operator between the two copies of the
$W_{1+\infty}$ algebra.  As abstract differential operators,
$X^{(k)}_n$ and $\Xbar^{(k)}_n$ commute with each other and
generate two independent $W_{1+\infty}$ algebras. As differential
operators acting on the $\tau$ function, however, they are related
(and this relation is nothing else but the string equations or the
$W_{1+\infty}$ constraints),
because the two vertex operators $X(z,w)$ and $\Xbar(z,w)$
correspond to the same fermion bilocal field $:\psi(z)\psi^*(w):$.
The only difference is whether the fermion bilocal field is inserted
to the left or right of $g$ in the fermionic representation of the
$\tau$ function. Thus, somewhat symbolically, the two $W_{1+\infty}$
algebras are related as
\beqn
    \{ \Xbar^{(k)}_n \} = g^{-1} \{ X^{(k)}_n \} g.
\eeqn
This is exactly what we observed throughout Section 3.

In fact, these two $W_{1+\infty}$ symmetries can also be viewed as
symmetries of two copies of the KP hierarchy. In fact, as already
mentioned, the $\tau$ function $\tau(t,\tbar,q)$ is also a $\tau$
function of the KP hierarchy with respect to both $t$ and $\tbar$.
In other words, two copies of the KP hierarchy are embedded into
the Toda lattice hierarchy.  Since the Lax formalism of the KP
hierarchy is formulated in terms of pseudo-differential operators,
one can consider two algebras of pseudo-differential operators
with two different spatial variables, say $X$ and $Y$, associated
with these copies of the KP hierarchy.  One can naturally obtain
two realizations of the $W_{1+\infty}$ algebra with generators
$X^k \rd_X^{n+k}$ and $Y^k \rd_Y^{n+k}$. In fact, following the
usual construction of the KP hierarchy, one can identify $X = t_1$
and $Y = \tbar_1$. This is a way how to justify the heuristic
interpretation of the two $W_{1+\infty}$ algebras in terms of
two kinds of pseudo-differential operators.

%%%%%%%%%%%%%%%%%%%%%%%%%%%%%%%%%%%%%%%%%%%%%%%%%%%%%%%%%%%%%%%%%%%%
\subsection{$c = 1$ strings at self-dual radius}

It is now rather straightforward to translate the contents of
Section 3 into the framework of the Toda lattice hierarchy.
The matrix $\bU$ is now diagonal,
\beqn
    \bU = \Bigl( e^{\alpha_i} \delta_{ij} \Bigr),
\eeqn
and satisfies the relations
\beqnarray
    \bU \Lambda^{-2} \bU^{-1}
    &=& \frac{1}{\mu^2 + m}
        \Bigl(
          - ( \Delta \Lambda^{-1} - i \mu \Lambda^{-1} )^2
          + m \Lambda^{-2}
        \Bigr),
                                               \nonumber \\
    \bU^{-1} \Lambda^2 \bU
    &=& \frac{1}{\mu^2 + m}
        \Bigl(
          - ( \Delta \Lambda + (1 - i \mu) \Lambda )^2
          + m \Lambda^2
        \Bigr)
\eeqnarray
as a consequence of  relations
(\ref{56}) and (\ref{75}) satisfied by the operator $g$.
These relations turn into the relations
\beqnarray
    \Lbar^{-2}
    &=& \frac{1}{\mu^2 + m}
        \Bigl( - ( M L^{-1} - i \mu L^{-1} )^2 + m L^{-2} \Bigr),
                                               \nonumber \\
    L^2
    &=& \frac{1}{\mu^2 + m}
        \Bigl( - ( \Mbar \Lbar + (1 - i \mu) \Lbar )^2
               + m \Lbar^2 \Bigr)
                                  \label{eq:Toda-StringEq}
\eeqnarray
among the Lax and Orlov-Shulman operators of the Toda lattice
hierarchy. In other words, they are (nonperturbative) string
equations of compactified $c = 1$ strings at self-dual radius.

The next issue is to give an interpretation of the classical limit
discussed in Section 3.  To this end, let us briefly review a standard
prescription of classical limit in the Toda lattice hierarchy
\cite{bib:TT-Toda}. The first step is to reformulate the Toda
lattice hierarchy in an $\hbar$-dependent way.  Naively, this is
achieved by rescaling the variables $(t,\tbar,q)$ as
\beqn
    t_n     \to \hbar^{-1} t_n, \quad
    \tbar_n \to \hbar^{-1} \tbar_n, \quad
    q       \to \hbar^{-1} q.
\eeqn
The lattice coordinate $q$ now takes values in the lattice $\hbar Z$,
i.e., we are considering a lattice with spacing $\hbar$. Actually,
this is just a heuristic argument; we rather restart from an
$\hbar$-dependent definition of the Lax-Orlov-Shulman operators,
\beqnarray
    L &=& e^{\hbar\rd_q} + \sum_{n=0}^\infty u_{n+1} e^{-n\hbar\rd_q},
                                                    \nonumber \\
    M &=& \sum_{n=1}^\infty n t_n L^n + q
        + \sum_{n=1}^\infty v_n L^{-n},
                                                    \nonumber \\
    \Lbar &=& \utilde_0 e^{\hbar\rd_q}
            + \sum_{n=0}^\infty \utilde_{n+1} e^{(n+2)\hbar\rd_q},
                                                    \nonumber \\
    \Mbar &=& - \sum_{n=1}^\infty n \tbar_n \Lbar^{-n} + q
            + \sum_{n=1}^\infty \vbar_n \Lbar^n,
\eeqnarray
and consider the $\hbar$-dependent twisted canonical commutation
relations
\beqn
   [ L , M ] = \hbar L, \quad [ \Lbar , \Mbar ] = \hbar \Lbar,
\eeqn
and Lax equations
\beqnarray
    \hbar \frac{\rd L}{\rd t_n} = [A_n, L], &\quad&
    \hbar \frac{\rd L}{\rd \tbar_n} = [\Abar_n, L],
                                              \nonumber \\
    \hbar \frac{\rd M}{\rd t_n} = [A_n, M], &\quad&
    \hbar \frac{\rd M}{\rd \tbar_n} = [\Abar_n, M],
                                              \nonumber \\
    \hbar \frac{\rd \Lbar}{\rd t_n} = [A_n, \Lbar], &\quad&
    \hbar \frac{\rd \Lbar}{\rd \tbar_n} = [\Abar_n, \Lbar],
                                              \nonumber \\
    \hbar \frac{\rd \Mbar}{\rd t_n} = [A_n, \Mbar], &\quad&
    \hbar \frac{\rd \Mbar}{\rd \tbar_n} = [\Abar_n, \Mbar].
\eeqnarray
$A_n$ and $\Abar_n$ are defined in the the same way as in the
case of $\hbar = 1$. (Note that we have also rescaled $M$ and
$\Mbar$ as
\beqn
    M     \to \hbar^{-1} M, \quad
    \Mbar \to \hbar^{-1} \Mbar,
\eeqn
so that they have the above expression.)  Upon this reformulation,
the coefficients $u_n$, etc. can depend on $\hbar$ in an arbitrary
way as $u_n = u_n(\hbar,t,\tbar,q)$, i.e., we do not have to assume
that they are obtained from $\hbar$-independent functions of
$(t,\tbar,q)$ by the above rescaling. To ensure the existence of
classical limit, however, we have to assume that
\beqnarray
    u_n(\hbar,t,\tbar,q) = u^{(0)}_n(t,\tbar,q) + O(\hbar), &&
    v_n(\hbar,t,\tbar,q) = v^{(0)}_n(t,\tbar,q) + O(\hbar),
                                              \nonumber \\
    u_n(\hbar,t,\tbar,q) = u^{(0)}_n(t,\tbar,q) + O(\hbar), &&
    v_n(\hbar,t,\tbar,q) = v^{(0)}_n(t,\tbar,q) + O(\hbar)
                                              \nonumber \\
&&
\eeqnarray
as $\hbar \to 0$, and construct from the leading terms the
following Laurent series of a new variable $P$:
\beqnarray
    \cL &=& P + \sum_{n=0}^\infty u^{(0)}_{n+1} P^{-n}
                                                    \nonumber \\
    \cM &=& \sum_{n=1}^\infty n t_n \cL^n + q
        + \sum_{n=1}^\infty v^{(0)}_n \cL^{-n},
                                                    \nonumber \\
    \cLbar &=& \utilde_0^{(0)} P
            + \sum_{n=0}^\infty \utilde^{(0)}_{n+1} P^{n+2},
                                                    \nonumber \\
    \cMbar &=& - \sum_{n=1}^\infty n \tbar_n \cLbar^{-n} + q
            + \sum_{n=1}^\infty \vbar^{(0)}_n \cLbar^n.
\eeqnarray
These Laurent series give a classical counterpart of the Lax
and Orlov-Shulman operators, and indeed turn out to obey the
classical twisted canonical relations
\beqn
    \{~\cL~,~ \cM~\} = \cL~,~ \quad \{~\cLbar~,~ \cMbar~\} = \cLbar
\eeqn
and the Lax equations
\beqnarray
    \frac{\rd \cL}{\rd t_n} = \{\cA_n, \cL\}, &\quad&
    \frac{\rd \cL}{\rd \tbar_n} = \{\cAbar_n, \cL\},
                                              \nonumber \\
    \frac{\rd \cM}{\rd t_n} = \{\cA_n, \cM\}, &\quad&
    \frac{\rd \cM}{\rd \tbar_n} = \{\cAbar_n, \cM\},
                                              \nonumber \\
    \frac{\rd \cLbar}{\rd t_n} = \{\cA_n, \cLbar\}, &\quad&
    \frac{\rd \cLbar}{\rd \tbar_n} = \{\cAbar_n, \cLbar\},
                                              \nonumber \\
    \frac{\rd \cMbar}{\rd t_n} = \{\cA_n, \cMbar\}, &\quad&
    \frac{\rd \cMbar}{\rd \tbar_n} = \{\cAbar_n, \cMbar\},
\eeqnarray
where $\{\quad,\quad\}$ denotes the new Poisson bracket
\beqn
    \{ A(P,q), B(P,q) \}
    = P \frac{\rd A(P,q)}{\rd P} \frac{\rd B(P,q)}{\rd q}
    - \frac{\rd A(P,q)}{\rd q} P \frac{\rd B(P,q)}{\rd P}.
\eeqn
This somewhat unusual definition is rather natural in view
of the quantum-classical correspondence
\beqn
    [~e^{\hbar \rd_q}~, ~q~] = \hbar e^{\hbar \rd_q} \
    \to \
    \{~ P~, ~q~ \} = P.
\eeqn
In fact, if we define $Q = qP^{-1}$, $P$ and $Q$ form a canonical
conjugate pair,
\beqn
    \{~ P~,~ Q~ \} = 1.
\eeqn
Thus this Poisson bracket is essentially the same as the Poisson
bracket of Section 3. The $W_{1+\infty}$ algebra of difference
operators turns into a $w_{1+\infty}$ algebra realized by this
Poisson bracket. Finally, $\cA_n$ and $\cAbar_n$ are given by
\beqn
    \cA_n = ( \cL^n )_{\ge 0}, \quad
    \cAbar_n = ( \cLbar^{-n} )_{<0},
\eeqn
where $(\quad)_{\ge 0}$ and $(\quad)_{<0}$ now stand for the
projection onto nonnegative and negative powers of $P$. This
classical limit of the Toda lattice hierarchy is also called the
``dispersionless Toda hierarchy'' in the literature
\cite{bib:Kodama}.

Let us return to the issue of $c = 1$ strings. We have derived
string equations (\ref{eq:Toda-StringEq}) in the $\hbar = 1$
formulation. An $\hbar$-dependent reformulation is obtained by
rescaling
\beqn
    q     \to \hbar^{-1} q, \quad
    M     \to \hbar^{-1} M, \quad
    \Mbar \to \hbar^{-1} \Mbar
\eeqn
along with similar rescaling of $t$ and $\tbar$. Actually, we have
seen in Section 3 that there are two different choices of $\hbar$
and, accordingly, two different prescriptions of classical limit.

In the first choice with $\hbar = - 1 /(i \sqrt{m})$, rescaling of
the string equations gives
\beqnarray
    \Lbar^{-2}
    &=& \frac{1}{\mu^2 - \hbar^{-2}}
        \Bigl( - ( \hbar^{-1} M L^{-1} - i \mu L^{-1} )^2
               - L^{-2} \Bigr),
                                               \nonumber \\
    L^2
    &=& \frac{1}{\mu^2 - \hbar^{-2}}
        \Bigl( - ( \hbar^{-1} \Mbar \Lbar + (1 - i \mu) \Lbar )^2
               - \Lbar^2 \Bigr).
\eeqnarray
In the classical limit as $\hbar \to 0$, these string equations are
replaced by
\beqnarray
    \cLbar^{-2} &=& \cM \cL^{-2} + \cL^{-2},
                                               \nonumber \\
    \cL^2       &=& \cMbar^2 \cLbar^2 + \cLbar^2.
\eeqnarray
This reproduces classical scattering relations (\ref{61})
upon identifying
\beqnarray
    P_{+} = \cLbar, &\quad&  Q_{+} = \cMbar \cLbar^{-1},
                                   \nonumber \\
    P_{-} = \cL,    &\quad&  Q_{-} = \cM \cL^{-1}.
\eeqnarray
Note that they are indeed a canonical pair with respect to the
Poisson bracket introduce above, i.e., $\{ P_\pm, Q_\pm \} = 1$.

Similarly, in the second choice with $\hbar = - 1/(i\mu)$,
we obtain the string equations
\beqnarray
    \Lbar^{-2}
    &=& \frac{1}{m - \hbar^{-2}}
        \Bigl( - (\hbar^{-1} M L^{-1} + \hbar^{-1} L^{-1})^2
               + m L^{-2} \Bigr),
                                             \nonumber \\
    L^2
    &=& \frac{1}{m - \hbar^{-2}}
        \Bigl( - (\hbar^{-1} \Mbar \Lbar
                  + (1 + \hbar^{-1}) \Lbar)^2
               + m \Lbar^2 \Bigr),
\eeqnarray
and their classical limit
\beqnarray
    \cLbar^{-2} &=& ( \cM \cL^{-1} + \cL^{-1} )^2,
                                               \nonumber \\
    \cL^2       &=& ( \cMbar \cLbar + \cLbar )^2.
\eeqnarray
These string equations, too, agree with  classical
scattering relations (\ref{58}) under
the same identification of $P_\pm$
and $Q_\pm$.

We have thus translated the scattering relations of Section 3
into string equations of the Lax-Orlov-Shulman operators of
the Toda lattice hierarchy. By identifying  $J_n$ in $H(t)$
and $J_{-n}$ in $\Hbar(t)$ with creation operators of massless
tachyons, the $\tau$ function can be  interpreted as
the partition function of $c = 1$ strings in the presence of
tachyon condensates besides the black hole background.
The time variables $t$ and $\tbar$ play the role of sources
that turn on tachyon condensation. The string equations can be
rewritten as $W_{1+\infty}$-constraints to the $\tau$ function.
Finding an explicit form of those constraints is now a rather
straightforward exercise.  This should however be a crucial
step for better understanding of higher orders of multi-loop
expansion.

%%%%%%%%%%%%%%%%%%%%%%%%%%%%%%%%%%%%%%%%%%%%%%%%%%%%%%%%%%%%%%%%%%
%%%%%%%%%%%%%%%%%%%%%%%%%%%%%%%%%%%%%%%%%%%%%%%%%%%%%%%%%%%%%%%%%%

\end{document}